\documentclass[reqno,12pt]{amsart}
\usepackage{fullpage}
\usepackage{amsfonts}
\usepackage{amssymb}

\numberwithin{equation}{section}
\def\s{}

\def\D{{\mathcal D}}
\def\H{{\mathcal H}}
\def\E{{\mathcal E}}
\def\R{{\mathcal R}}
\def\K{{\mathcal K}}

\def\I{{\mathcal I}}

\def\inv{^{-1}}
\def\X{{\rm X}}
\def\pr{{\rm pr}}
\def\lieder{{\mathfrak L}}
\def\sgn{{\rm sgn}}

\def\Re{{\rm Re}\;}
\def\Im{{\rm Im}\;}
\def\inum{{\rm i}}

\def\Rnum{\mathbb{R}}
\def\Znum{\mathbb{Z}}

\def\wtil{\widetilde}
\def\what{\widehat}
\def\txtbinom#1#2{{\textstyle\binom{#1}{#2}}}

\def\const{{\rm const.}}
\def\Rnum{\mathbb{R}}
\def\Cnum{\mathbb{C}}
\def\Re{{\rm Re}\,}
\def\Im{{\rm Im}\,}
\def\scal{{\rm scal.}}

\def\w{w}
\def\k{k}

\def\Ref#1{Ref.\cite{#1}}

\newtheorem{lem}{Lemma}

\def\scrpt#1{$\scriptstyle {#1}$}

\begin{document}
\allowdisplaybreaks[3]

\title{Integrable U(1)-invariant peakon equations\\ from the NLS hierarchy}

\author{
Stephen C. Anco${}^1$
\\\lowercase{\scshape{ and }}\\
Fatane Mobasheramini${}^{1,2}$
\\
\\\lowercase{\scshape{
${}^1$
Department of Mathematics and Statistics\\
Brock University\\
St. Catharines, ON L\scrpt2S\scrpt3A\scrpt1, Canada}} \\
\\\lowercase{\scshape{
${}^2$
Department of Mathematics and Statistics\\
Concordia University\\
Montreal, QC H\scrpt3G\scrpt1M\scrpt8, Canada}}
}

\thanks{sanco@brocku.ca, fatane.mobasheramini@mail.concordia.ca.}

\begin{abstract}
Two integrable $U(1)$-invariant peakon equations 
are derived from the NLS hierarchy through the tri-Hamiltonian splitting method.
A Lax pair, a recursion operator, a bi-Hamiltonian formulation, 
and a hierarchy of symmetries and conservation laws
are obtained for both peakon equations. 
These equations are also shown to arise as potential flows in the NLS hierarchy
by applying the NLS recursion operator to flows generated by 
space translations and $U(1)$-phase rotations on a potential variable. 
Solutions for both equations are derived 
using a peakon ansatz combined with an oscillatory temporal phase. 
This yields the first known example of a peakon breather. 
Spatially periodic counterparts of these solutions are also obtained. 
\end{abstract}

\maketitle

\section{Introduction}
\label{intro}

There has been considerable recent interest in equations that possess 
peaked solitary wave solutions, known as peakons. 
One of the first well-studied peakon equations is the Camassa-Holm (CH) equation 
\cite{CamHol,CamHolHym} 
$u_t -u_{txx} + 3u u_x - 2 u_x u_{xx} - u u_{xxx}=0$, 
which arises from the theory of shallow water waves. 
This equation possesses peakon solutions, $u=c \exp(-|x-ct|)$, 
and multi-peakon solutions which are linear superpositions of single peakons 
with time-dependent amplitudes and speeds. 
The CH equation also possesses a large class of solutions 
in which smooth initial data evolves to form a cusped wave in a finite time 
(i.e., a gradient blow up) while the wave amplitude remains bounded.
These features are shared by many other equations, 
all of which belong to a general family of peakon equations \cite{AncRec}
$u_t -u_{txx} + f(u,u_x)(u-u_{xx}) + (g(u,u_x)(u-u_{xx}))_x=0$
where $f(u,u_x)$ and $g(u,u_x)$ are arbitrary non-singular functions. 

The CH equation is also an integrable system \cite{CamHol,FisSch,FucFok,Qia03}. 
In particular, it has a Lax pair, a bi-Hamiltonian formulation,
and an infinite hierarchy of symmetries and conservation laws 
generated by a recursion operator. 
Moreover, 
the CH equation is related to the integrable hierarchy that contains the Korteweg-de Vries (KdV) equation
$v_t + vv_x +v_{xxx}=0$, 
which itself is an integrable system arising from the theory of shallow water waves.
Firstly, 
the CH equation can be obtained from a negative flow in the KdV hierarchy 
by a hodograph transformation \cite{Fuc}. 
Secondly, 
the CH equation also can be derived as a potential flow 
by applying the KdV recursion operator to the flow generated by $u_x$ 
where $v=u-u_{xx}$ relates the variables in the two equations \cite{ChoQu}. 
Thirdly, 
the Hamiltonian structures of the CH equation can be derived 
from those of the KdV equation by a tri-Hamiltonian splitting method \cite{OlvRos}. 

The KdV equation is well-known to be related to the modified KdV (mKdV) equation
$v_t + v^2 v_x +v_{xxx}=0$
by a Miura transformation. 
This equation is also a well-known integrable system, 
and it is related to a peakon equation 
$u_t -u_{txx} + ((u^2- u_x^2)(u -u_{xx}))_x =0$
called the modified CH (mCH) equation 
(also known as the Fokas-Olver-Rosenau-Qiao (FORQ) equation),
which is an integrable system. 
In particular, 
the mCH equation arises from the theory of surface water waves \cite{Fok95b}
and its integrability was derived by applying the tri-Hamiltonian splitting method
to the Hamiltonian structures of the mKdV equation \cite{OlvRos,Fok95a,FokOlvRos}. 
It can also be obtained as a potential flow by using the mKdV recursion operator. 
A derivation based on spectral methods is given in \Ref{Qia06} 
where the Lax pair and single peakon solutions of the mCH equation were first obtained. 
Other recent work on the mCH equation appears in \Ref{Qia07,QiaLi,GuiLiuOlvQu,KanLiuOlvQu}.

The main purpose of the present paper is 
to derive complex, $U(1)$-invariant peakon equations
from the integrable hierarchy that contains the nonlinear Schrodinger (NLS) equation. 
Two different peakon equations will be obtained. 
One is a complex analog of the mCH equation. 
The other is a peakon analog of the NLS equation itself. 
Both of these peakon equations are integrable systems,
and the main aspects of their integrability will be presented:
a Lax pair, a bi-Hamiltonian formulation, a recursion operator,
and an infinite hierarchy of symmetries and conservation laws. 

These two peakon equations were first derived as 2-component coupled systems 
in \Ref{XiaQia,XiaQiaZho}
by Lax pair methods, without consideration of the NLS hierarchy. 
The derivation in the present paper shows how both of these two equations 
describe negative flows in the NLS hierarchy
and provides a common origin for their Lax pairs. 
In addition, their integrability structure is obtained in a simple, systematic way 
by combining the tri-Hamiltonian splitting method 
and the AKNS zero-curvature method \cite{AblKauNewSeg}. 

In section~\ref{method},
the derivation of the mCH equation from the mKdV hierarchy will be briefly reviewed,
using the tri-Hamiltonian splitting method, 
which utilizes the bi-Hamiltonian structure of the mKdV hierarchy, 
and using the recursion operator method,
which is closely connected to a zero-curvature matrix formulation of the mKdV hierarchy. 
Compared to the original presentations of these two methods 
\cite{Fuc,OlvRos,FokOlvRos} and \cite{ChoQu}, 
some new aspects will be developed, 
including an explicit recursion formula for all of the Hamiltonians in the mCH hierarchy,
and a derivation of the mCH recursion operator directly from 
the AKNS zero-curvature equation \cite{AblKauNewSeg}. 

In section~\ref{splitting}, 
the tri-Hamiltonian splitting method will be applied to the Hamiltonian structures 
in the NLS hierarchy.
Rather than use the standard bi-Hamiltonian structure of the NLS equation itself, 
whose tri-Hamiltonian splitting is known to lead to a somewhat trivial equation \cite{OlvRos}, 
a third Hamiltonian structure of the NLS equation \cite{MarSanWan,AncMyr}
will be used instead. 
This Hamiltonian structure is connected with a higher flow in the NLS hierarchy, 
given by the Hirota equation \cite{Hir1973}, 
which is a complex, $U(1)$-invariant generalization of the mKdV equation. 

In section~\ref{derivation}, 
the Hamiltonian operators derived from the tri-Hamiltonian splitting method
will be used to construct the two $U(1)$-invariant peakon equations
along with their bi-Hamiltonian formulation. 
These two peakon equations will also be shown to arise 
as potential flows in the NLS hierarchy
by applying the NLS recursion operator to the flows generated by 
$x$-translations and $U(1)$-phase rotations on a potential variable. 
One of the peakon equations has an NLS-type form 
which does not admit a real reduction,
while the other peakon equation has a complex mCH-type form
whose real reduction is given by the mCH equation. 

In section~\ref{laxpair},
for each of the two $U(1)$-invariant peakon equations, 
their bi-Hamiltonian structure will be used to derive 
a hierarchy of symmetries and conservation laws, 
and a Lax pair will be obtained by modifying 
the zero-curvature equation of the mCH equation. 

In section~\ref{peakons}, 
solutions for the two $U(1)$-invariant peakon equations will be derived
by using a peaked travelling wave expression modified by a temporal phase oscillation, 
$u=a \exp(\inum(\phi+\omega t)-|x-ct|)$. 
Specifically, 
the complex mCH-type equation will be shown to possess only 
a standard peakon with a constant phase, 
while in contrast the NLS-type peakon equation will be shown to possess 
a stationary peakon with a temporal phase oscillation given by 
$w=\tfrac{1}{3}a^3$. 
This provides the first ever example of a peakon breather. 
Moreover, spatially periodic counterparts of these peakons will be derived. 

Finally, some concluding remarks will be made in section~\ref{remarks}.

\section{Derivation of the mCH peakon equation}
\label{method}

We start by considering the mKdV equation 
\begin{equation}\label{mkdv-eqn}
v_t + \tfrac{3}{2}v^2 v_x +v_{xxx} =0
\end{equation}
where we have chosen the scaling factor in the nonlinear term 
to simplify subsequent expressions. 
This equation has the bi-Hamiltonian structure
\begin{equation}\label{mkdv-biHam-eqn}
v_t=\H(\delta H/\delta v)=\E(\delta E/\delta v)
\end{equation}
where
\begin{align}
\H & =-D_x , 
\label{mkdv-Hop}
\\
\E & =-D_x^3-D_xvD\inv_x vD_x
\label{mkdv-Eop}
\end{align}
are compatible Hamiltonian operators,
and where
\begin{align}
H & =\int_{-\infty}^{\infty} \tfrac{1}{8}v^4 -\tfrac{1}{2}v_x^2\,dx , 
\\
E & =\int_{-\infty}^{\infty} \tfrac{1}{2}v^2\,dx
\end{align}
are the corresponding Hamiltonians. 

Recall \cite{Olv-book}, 
a linear operator $\D$ is a Hamiltonian operator iff 
its associated bracket 
\begin{equation}\label{bracket}
\{H,E\}_\D = \int_{-\infty}^{\infty} (\delta H/\delta v)\D(\delta E/\delta v)\, dx
\end{equation}
is skew and satisfies the Jacobi identity,
for all functionals $H$ and $E$. 
Two Hamiltonian operators are called compatible if every linear combination of them 
is a Hamiltonian operator.
Also recall, 
the composition of a Hamiltonian operator and the inverse of a compatible Hamiltonian operator is a recursion operator that obeys a hereditary property \cite{FucFok}. 
In particular, a linear operator $\R$ is hereditary if it satisfies 
\begin{equation}\label{hereditaryprop}
\lieder_{\X_{\R\eta}}\R = \R(\lieder_{\X_{\eta}}\R)
\end{equation}
holding for all differential functions $\eta(x,v,v_x,\ldots)$,
where $\lieder_{\X_f}$ denotes the Lie derivative with respect to a vector field 
$\X_f= f(x,v,v_x,\ldots)\partial_v$. 
An equivalent formulation of the hereditary property is given by \cite{Olv-book}
\begin{equation}\label{hereditaryop}
\pr\X_{\R\eta}\R +[\R,(\R\eta){}^\prime]= \R(\pr\X_{\eta}\R+[\R,\eta{}^\prime]) 
\end{equation}
where $f^\prime$ denotes the Frechet derivative of 
a differential function $f(x,v,v_x,\ldots)$
and $\pr\X_f$ denotes the prolongation of $\X_f$ to the coordinate space $J^\infty=(x,v,v_x,\ldots)$. 

The compatible Hamiltonian operators \eqref{mkdv-Hop}--\eqref{mkdv-Eop} 
yield the hereditary recursion operator
\begin{equation}\label{mkdv-Rop}
\R=\E\H\inv=D_x^2+v^2 +v_xD\inv_x v . 
\end{equation}
This operator and both of the Hamiltonian operators 
are invariant under $x$-translations applied to $v$,
which corresponds to the invariance of the mKdV equation under 
the symmetry operator $\X=-v_x\partial/\partial_v$
representing infinitesimal $x$-translations. 
The recursion operator can be combined with this symmetry to express 
the mKdV equation as a flow
\begin{equation}
v_t=\R({-v_x})=-\tfrac{3}{2}v^2 v_x -v_{xxx} . 
\end{equation}
Higher flows are generated by 
\begin{equation}\label{mkdv-hierarchy}
P^{(n)}=\R^n({-v_x}),
\quad
n=1,2,\ldots
\end{equation}
corresponding to an integrable hierarchy of equations $v_t= P^{(n)}$,
where the $n=1$ flow is the mKdV equation
and each successive flow $n\geq 2$ inherits a bi-Hamiltonian structure 
$P^{(n)} = \H(\delta H^{(n)}/\delta v)=\E(\delta H^{(n-1)}/\delta v)$
which comes from Magri's theorem,
with $H^{(0)}=E$ and $H^{(1)}=H$. 
The gradients of these Hamiltonians are generated by 
$\delta H^{(n)}/\delta v = {\R^*}^n(v) = Q^{(n)}$, $n=1,2,\ldots$, 
in terms of the adjoint recursion operator 
\begin{equation}\label{mkdv-adRop}
\R^*=\H\inv\E=D_x^2+v^2 -vD\inv_x v_x
\end{equation}
where $Q^{(0)}=\delta H^{(0)}/\delta v=\delta E/\delta v = v$. 

It is useful to observe that $P^{(0)}=-v_x$ also can be expressed as 
a bi-Hamiltonian flow, in a certain formal sense. 
Its first Hamiltonian structure is simply 
$P^{(0)} = \H(\delta H^{(0)}/\delta v)$, 
while its second Hamiltonian structure arises from the relation 
$\E(0) = -v_xD\inv_x(0)$ 
if $D\inv_x(0)$ is redefined by the addition of some non-zero constant,
so that $D\inv_x(0)=c\neq 0$. 
This yields $P^{(0)} = c^{-1}\E(0)$, 
where the corresponding Hamiltonian is trivial. 
Then this flow has the bi-Hamiltonian structure
\begin{equation}\label{mkdv-rootflow}
P^{(0)} = \H(\delta H^{(0)}/\delta v) =c^{-1}\E(0) . 
\end{equation}

Each higher flow in the mKdV hierarchy \eqref{mkdv-hierarchy}
corresponds to a higher symmetry operator 
$\X^{(n)}=P^{(n)}\partial/\partial_v$, $n=1,2,\ldots$, 
all of which represent infinitesimal symmetries that are admitted by 
every equation $v_t= P^{(n)}$,  $n=1,2,\ldots$, in the hierarchy. 
Each Hamiltonian $H^{(n)}$, $n=0,1,2,\ldots$, of the flows in the hierarchy
corresponds to a conservation law $\frac{d}{dt} H^{(n)}=0$, 
all of which hold for every equation $v_t= P^{(n)}$,  $n=1,2,\ldots$, in the hierarchy. 

The mKdV equation has the scaling symmetry 
$v\to\lambda v$, $x\to\lambda^{-1}x$, $t\to\lambda^{-3}t$. 
This symmetry can be used to derive a simple scaling formula that yields 
the Hamiltonians $H^{(n)}$ in the mKdV hierarchy, 
by the general scaling method \cite{Anc03} shown in the appendix. 
The formula is given by 
\begin{equation}\label{mkdv-H-Q-expr}
H^{(n)}=\int_{-\infty}^{\infty}h^{(n)}\,dx, 
\quad
h^{(n)} = \tfrac{1}{2n+1} D\inv_x(vD_xQ^{(n)}) 
\end{equation}
where the Hamiltonian density $h^{(n)}$ can be freely changed by the addition of 
a total $x$-derivative. 
Furthermore, 
the gradient recursion formula $Q^{(n)}= {\R^*}^n(v)$ 
can be used to convert the Hamiltonian expression \eqref{mkdv-H-Q-expr}
into a recursion formula for $h^{(n)}$ itself, as follows.
First, note expression \eqref{mkdv-H-Q-expr} can be inverted to give 
$Q^{(n)}= (2n+1) D\inv_xv^{-1} D_x h^{(n)}$. 
Next, replacing $Q^{(n)}= {\R^*}Q^{(n-1)} =(2n-1) (D_xv^{-1} D_x h^{(n-1)}+v h^{(n-1)})$
in expression \eqref{mkdv-H-Q-expr} yields
$(2n+1)h^{(n)} = (2n-1)D\inv_x\big(vD_x(v+D_x v^{-1}D_x) h^{(n-1)}\big)$
modulo a total $x$-derivative, 
and thus
\begin{equation}\label{mkdv-H-expr}
h^{(n)} = \tfrac{2n-1}{2n+1}D\inv_x\big(vD_xv(1+(v^{-1}D_x)^2)h^{(n-1)}\big) . 
\end{equation}
This provides an explicit recursion formula for all of the Hamiltonians in the mKdV hierarchy, 
starting from the Hamiltonian density $h^{(0)}= \tfrac{1}{2}v^2$. 

\subsection{Tri-Hamiltonian splitting method}

The basis for the method of tri-Hamiltonian splitting \cite{OlvRos} is the observation that 
if a linear operator $\D$ is Hamiltonian then so is the scaled operator 
$\D_{(\lambda)}$ defined by scaling $v\to \lambda v$ 
(and similarly scaling all $x$-derivatives of $v$) in $\D$. 
When $\D_{(\lambda)}$ consists of two terms with different powers of $\lambda$,
each term will define a Hamiltonian operator, 
and the resulting two operators can be shown to be a compatible Hamiltonian pair. 

Under scaling, 
the first mKdV Hamiltonian operator \eqref{mkdv-Hop} is invariant, 
$\H_{(\lambda)} = -D_x = \H$, 
while the second mKdV Hamiltonian operator \eqref{mkdv-Eop} becomes 
$\E_{(\lambda)} = -D_x^3 -\lambda^2 D_xvD\inv_x vD_x$,
which yields two operators 
\begin{equation}\label{mkdv-splitEops}
\E_1=-D_x^3,
\quad
\E_2=-D_xvD\inv_x vD_x . 
\end{equation}
These operators \eqref{mkdv-splitEops} are a compatible Hamiltonian pair. 
Furthermore, 
since the first operator \eqref{mkdv-Hop} is invariant, 
all three operators \eqref{mkdv-splitEops} and \eqref{mkdv-Hop} 
can be shown to be mutually compatible.
Hence, 
$\H$, $\E_1$, $\E_2$ constitute a compatible Hamiltonian triple. 
These operators can be combined to obtain a new pair of compatible Hamiltonian operators:
\begin{equation}\label{mkdv-peakon-ops}
\wtil\H = \H - \E_1=-D_x + D_x^3, 
\quad
\wtil\E= \E_2=-D_xvD\inv_x vD_x . 
\end{equation}

The following steps are used to derive a peakon equation from this Hamiltonian pair \eqref{mkdv-peakon-ops}. 
First, observe that the operator $\wtil\H$ has the factorization 
\begin{equation}\label{mkdv-Delaops}
\wtil\H = -D_x\Delta = -\Delta D_x,
\quad
\Delta = 1- D_x^2
\end{equation}
where $\Delta=\Delta^*$ is a symmetric operator. 
Second, introduce a potential variable $u$ in terms of $v$, 
\begin{equation}\label{mkdv-potential}
v = \Delta u
\end{equation}
where these variables satisfy the variational relation 
\begin{equation}\label{mkdv-varders}
\Delta\frac{\delta}{\delta v} =\frac{\delta}{\delta u} . 
\end{equation}
Next, consider the flow defined by $x$-translations on $v$, 
$P^{(0)} = -v_x$. 
This root flow inherits a natural bi-Hamiltonian structure 
with respect to the two Hamiltonian operators \eqref{mkdv-peakon-ops}. 
The first Hamiltonian structure arises from expressing 
$-v_x = \wtil\H(\Delta\inv v)$ with $\Delta\inv v= \delta \wtil H/\delta v$, 
so then 
\begin{equation}
-v_x = \wtil\H(\delta \wtil H/\delta v)
\end{equation}
with $v= \Delta(\delta \wtil H/\delta v) = \delta \wtil H/\delta u$
through the variational relation \eqref{mkdv-varders}, 
where the Hamiltonian is given by 
\begin{equation}\label{mkdv-peakon-H}
\wtil H = \int_{-\infty}^{\infty} \tfrac{1}{2} uv\,dx
= \int_{-\infty}^{\infty} \tfrac{1}{2}(u^2 +u_x^2)\,dx
\end{equation}
using expression \eqref{mkdv-potential} and integrating by parts. 
The second Hamiltonian structure comes directly from splitting 
the operator $\E=\E_1+\E_2$ in the bi-Hamiltonian equation \eqref{mkdv-rootflow}
for $P^{(0)} = -v_x$. 
This yields 
$c^{-1}\E(0)= c^{-1}\E_2(0) = -v_x$,
since $\E_1(0) =0$. 
Hence, 
\begin{equation}
-v_x = c^{-1}\E(0)
\end{equation}
holds with a trivial Hamiltonian. 

Then, Magri's theorem \cite{Mag} implies that the bi-Hamiltonian flow 
\begin{equation}
P^{(0)} = -v_x = \wtil\H(\delta \wtil H/\delta v) = c^{-1}\wtil\E(0)
\end{equation}
defined by $x$-translations 
is a root flow for an integrable hierarchy of higher flows that are generated by 
the hereditary recursion operator 
\begin{equation}\label{mkdv-peakon-Rop}
\wtil\R=\wtil\E\wtil\H\inv= D_x v D\inv_x v\Delta\inv . 
\end{equation}
Each flow in the resulting hierarchy 
$\wtil P^{(n)} = \wtil\R^n({-v_x})$, $n=1,2,\ldots$, 
yields an equation $v_t = \wtil P^{(n)}$ that has a bi-Hamiltonian structure
$\wtil P^{(n)} = \wtil\H(\delta \wtil H^{(n)}/\delta v)=\wtil\E(\delta \wtil H^{(n-1)}/\delta v)$
with $\wtil H^{(0)}=\wtil H$. 
The gradients of the Hamiltonians $\wtil H^{(n)}$, $n=0,1,2,\ldots$, 
are generated by 
$\delta \wtil H^{(n)}/\delta v = \wtil\R^*{}^n(u) =\wtil Q^{(n)}$, $n=1,2,\ldots$, 
in terms of the adjoint recursion operator 
\begin{equation}\label{mkdv-peakon-adRop}
\wtil\R^*=\wtil\H\inv\wtil\E=\Delta\inv v D\inv_x vD_x
\end{equation}
where $\delta \wtil H^{(0)}/\delta v=\delta \wtil H/\delta v = u$. 
These Hamiltonian gradients are equivalently given by 
\begin{equation}\label{mkdv-peakon-Q}
\delta \wtil H^{(n)}/\delta u = \wtil\K^n(v) = \Delta \wtil Q^{(n)}
\end{equation}
in terms of the operator 
\begin{equation}\label{mkdv-peakon-Kop}
\wtil\K=\Delta\wtil\R^*\Delta\inv=v D\inv_x vD_x\Delta\inv . 
\end{equation}
Moreover, as $\wtil\K$ is scaling homogeneous in terms of $u$, 
the Hamiltonians can be obtained by the general scaling method \cite{Anc03}
shown in the appendix. 
This yields the formula 
\begin{equation}\label{mkdv-peakon-H-Q-expr}
\wtil H^{(n)} = \int_{-\infty}^{\infty} \tilde h^{(n)}\,dx,
\quad
\tilde h^{(n)}= \tfrac{1}{2(n+1)} u \wtil Q^{(n)}
\end{equation}
modulo boundary terms. 
Note this formula reproduces the Hamiltonian \eqref{mkdv-peakon-H} for $n=0$. 

Finally, the $n=1$ flow 
\begin{equation}\label{mch-flow}
\wtil P^{(1)} = \wtil\R({-v_x}) = - \tfrac{1}{2} ((u^2-u_x^2)v)_x
\end{equation}
yields the mCH equation 
\begin{equation}\label{mch-eqn}
v_t = - \tfrac{1}{2} ((u^2-u_x^2)v)_x 
= \wtil\H(\delta \wtil E/\delta v)=\wtil\E(\delta \wtil H/\delta v)
\end{equation}
along with its bi-Hamiltonian structure,
where $\wtil H=\wtil H^{(0)}$ is the Hamiltonian \eqref{mkdv-peakon-H}
and $\wtil E=\wtil H^{(1)}$ is the Hamiltonian given by 
\begin{equation}\label{mkdv-peakon-E}
\wtil E = \int_{-\infty}^{\infty} \tfrac{1}{8}( u(u^2-u_x^2)v ) \,dx
= \int_{-\infty}^{\infty} \tfrac{1}{24}( 3u^4 +6 u^2 u_x^2 -u_x^4) \,dx
\end{equation}
from formula \eqref{mkdv-peakon-H-Q-expr} with $n=1$, 
after integration by parts. 
Thus, we will refer to the integrable hierarchy of higher flows 
\begin{equation}\label{mch-hierarchy}
\wtil P^{(n)} = \wtil\R^n({-v_x}),
\quad
n=1,2,\ldots
\end{equation} 
as the mCH hierarchy. 

The flows in the mCH hierarchy correspond to symmetry operators
$\X^{(n)}=\wtil P^{(n)}\partial/\partial_v$, $n=1,2,\ldots$, 
while the Hamiltonians $\wtil H^{(n)}$ for these flows 
correspond to conservation laws $\frac{d}{dt}\wtil H^{(n)}=0$, $n=0,1,2,\ldots$,
all of which are admitted by each equation 
$v_t= \wtil P^{(n)}$,  $n=1,2,\ldots$, in the hierarchy. 

A useful remark is that 
the recursion formula \eqref{mkdv-peakon-Q} for Hamiltonian gradients
can be used to convert the Hamiltonian expression \eqref{mkdv-peakon-H-Q-expr}
into a recursion formula for the Hamiltonian densities, 
similarly to the mKdV case. 
Specifically, 
substitution of $\wtil Q^{(n)}= \K \wtil Q^{(n-1)} =2n u^{-1} \tilde h^{(n-1)}$
into expression \eqref{mkdv-peakon-H-Q-expr} gives 
\begin{equation}\label{mch-H-expr}
\tilde h^{(n)} =\tfrac{n}{n+1} uvD\inv_x\big(vD_x\Delta\inv(u^{-1}\tilde h^{(n-1)})\big)
\end{equation}
modulo a total $x$-derivative. 
This provides an explicit recursion formula for all of the Hamiltonians in the mCH hierarchy, 
starting from the Hamiltonian density $\tilde h^{(0)}= \tfrac{1}{2}vu$. 

All of the higher symmetries and higher Hamiltonian densities 
are found to be nonlocal expressions in terms of $u$, $u_x$, $v$, and $x$-derivatives of $v$. 
In particular, 
the first higher symmetry $\X^{(2)}=\wtil P^{(2)}\partial/\partial_v$
admitted by the mCH equation is given by 
\begin{equation}
\wtil P^{(2)} = \wtil\R\wtil P^{(1)} 
= \tfrac{1}{2} \big(v(uw-u_xw_x - \tfrac{1}{4}(u^2-u_x^2)^2)\big)_x
\end{equation}
where $w$ is another potential which is defined by 
\begin{equation}
\Delta w = v(u^2-u_x^2) . 
\end{equation}
Likewise, 
the first higher Hamiltonian density $\tilde h^{(2)}$ 
admitted by the mCH equation is given by 
\begin{equation}
\tilde h^{(2)} =\tfrac{2}{3} uvD\inv_x\big(vD_x\Delta\inv(u^{-1}\tilde h^{(1)})\big)
= \tfrac{1}{12}vu(uw -u_xw_x-\tfrac{1}{4}(u^2-u_x^2)^2)
\end{equation}
which yields the Hamiltonian 
\begin{equation}
\begin{aligned}
\tilde H^{(2)} & = \int_{-\infty}^{\infty} \tfrac{1}{12}vu(uw -u_xw_x-\tfrac{1}{4}(u^2-u_x^2)^2)\, dx
\\
& = \int_{-\infty}^{\infty} \tfrac{1}{12}( u^3w +\tfrac{3}{2} uu_x^2w - \tfrac{1}{2} u_x^3 w_x +\tfrac{1}{4}u^2u_x^2(u^2+u_x^2) -\tfrac{1}{4} u^6 -\tfrac{1}{10}u_x^6  )\, dx
\end{aligned}
\end{equation}
modulo boundary terms. 

\subsection{Recursion operator method}

From the splitting of the mKdV Hamiltonian operators, 
the recursion operator \eqref{mkdv-peakon-Rop} in the mCH hierarchy 
has a simple relationship to the mKdV recursion operator \eqref{mkdv-Rop}. 
First, 
the split operators \eqref{mkdv-splitEops} can be expressed in terms of 
the mCH operators \eqref{mkdv-peakon-ops} by 
\begin{equation}
\H=\Delta\inv\wtil\H,
\quad
\E_1 = (1-\Delta) \H, 
\quad
\E_2 =\wtil\E . 
\end{equation} 
Next, substitution of these operator expressions into the mKdV recursion operator \eqref{mkdv-Rop}
yields
\begin{equation}\label{mkdv-Rrel1}
\R=\E\H\inv= \E_1\H\inv +\E_2\H\inv
\end{equation}
where
\begin{equation}\label{mkdv-Rrel2}
\E_1\H\inv = 1-\Delta, 
\quad
\wtil\E_2\H\inv = \wtil\E_2\wtil\H\inv\Delta = \wtil\R\Delta . 
\end{equation}
Then, when these two equations \eqref{mkdv-Rrel1}--\eqref{mkdv-Rrel2} are combined, 
this gives the relation
\begin{equation}\label{mkdv-recursion-rel}
\R-1 =(\wtil\R-1)\Delta . 
\end{equation}

Since $\wtil\R$ is a hereditary recursion operator, 
so is $\wtil\R-1$, and therefore
\begin{equation}\label{mch-shifted-Rop}
\what\R = (\R-1)\Delta\inv
\end{equation}
defines a hereditary recursion operator. 
The hereditary property \eqref{hereditaryprop} of this operator 
can be verified to hold directly, 
using the hereditary property of the mKdV recursion operator $\R$,
which does not require splitting of the mKdV Hamiltonian operators. 

Furthermore, since $\R$ is invariant under $x$-translations, 
so is $\what\R$. 
Therefore, by a standard result in the theory of recursion operators 
\cite{Olv77,Olv-book}, 
$\what\R$ can be used to generate an integrable hierarchy of flows
$\what P^{(n)} = \what\R^n(-v_x)$, $n=1,2,\ldots$, 
starting from the root flow $P^{(0)}=-v_x$ for the mKdV and mCH hierarchies. 
Note the integrability structure provided by the recursion operator $\what\R$ 
for these flows consists of a hierarchy of higher symmetries 
$\X^{(n)}=\what P^{(n)}\partial_v$, $n=1,2,\ldots$, 
that are admitted by every equation $v_t= \what P^{(n)}$,  $n=1,2,\ldots$, 
in the hierarchy. 

The first flow in the resulting hierarchy is given by
a linear combination of the root flow and the first flow 
in the mCH hierarchy \eqref{mch-hierarchy}:
$\hat P^{(1)} = \wtil P^{(1)} - P^{(0)} = \what\R(-v_x)  = (\R-1)(-u_x)$
where 
\begin{equation}
\R(-u_x) = -u_{xxx} - \tfrac{1}{2}((u^2-u_x^2)v)_x . 
\end{equation}
This yields
\begin{equation}\label{mch-variant-flow}
\hat P^{(1)} = \wtil P^{(1)} - P^{(0)} = (\R-1)(-u_x) = v_x - \tfrac{1}{2}((u^2-u_x^2)v)_x . 
\end{equation}
Since all flows in that hierarchy have a bi-Hamiltonian structure 
given by the pair of mCH Hamiltonian operators \eqref{mkdv-peakon-ops}, 
the flow \eqref{mch-variant-flow} is also bi-Hamiltonian. 
The corresponding bi-Hamiltonian equation $v_t=\hat P^{(1)}$ 
is a slight generalization of the mCH equation \eqref{mch-eqn}:
\begin{equation}\label{mch-variant-eqn}
v_t - v_x + \tfrac{1}{2} ((u^2-u_x^2)v)_x =0 . 
\end{equation}
In particular, 
a Galilean transformation $x\rightarrow \tilde x=x+t$, $t\rightarrow\tilde t=t$
applied to the mCH equation \eqref{mch-eqn} 
yields the generalized equation \eqref{mch-variant-eqn}. 

\subsection{Lax pair}

The shifted recursion operator \eqref{mch-shifted-Rop}, 
which generates linear combinations of flows in the mCH hierarchy \eqref{mch-hierarchy}, 
is closely connected to a Lax pair for the mCH equation \eqref{mch-eqn}. 

To explain the connection, 
we first consider the mKdV shifted recursion operator $\R-1$. 
This operator arises in the following way from the AKNS scheme \cite{AblKauNewSeg}
using a matrix zero-curvature equation 
\begin{equation}\label{zero-curv}
U_t-V_x +[U,V]=0
\end{equation}
given by the matrices 
\begin{align}
U & = 
\begin{pmatrix}
\lambda & \tfrac{1}{2} v
\\
-\tfrac{1}{2} v & -\lambda
\end{pmatrix}, 
\label{real-akns-U}
\\
V & = 
\begin{pmatrix}
\lambda K & \tfrac{1}{2} \omega + \lambda P
\\
-\tfrac{1}{2} \omega + \lambda P & -\lambda K 
\end{pmatrix}
\label{real-akns-V}
\end{align}
where $\lambda$ is the spectral parameter. 
These matrices belong to the Lie algebra $sl(2,\Rnum)$, 
with $U$ being a reduction of the standard AKNS matrix to the case of 
a real spectral parameter $\lambda$ and a real variable $v$, 
and with $V$ being parameterized such that 
$\begin{pmatrix} K & 0 \\ 0 & -K \end{pmatrix}$,
$\begin{pmatrix} 0 & \omega \\ -\omega & 0 \end{pmatrix}$, 
$\begin{pmatrix} 0 & P \\ P & 0 \end{pmatrix}$
are mutually orthogonal in the Cartan-Killing inner product 
(as defined by the trace of a matrix product) 
in $sl(2,\Rnum)$.
The components of the zero-curvature equation \eqref{zero-curv} yield
\begin{subequations}
\begin{gather}
v_t = D_x \omega -4\lambda^2 P , 
\label{real-akns-flow-eqn}
\\
\omega = v K + D_x P , 
\label{real-akns-grad-eqn}
\\
D_x K = v P . 
\label{real-akns-aux-eqn}
\end{gather}
\end{subequations}
The second and third equations can be used to express 
$K=D\inv_x(vP)$ and $\omega = v D\inv_x(vP) + D_x P$,
and then the first equation becomes 
$v_t = D_x^2P +D_x(vD\inv_x(vP)) -4\lambda^2 P$. 
If we now put $\lambda =\tfrac{1}{2}$ then this yields
\begin{equation}\label{mkdv-lax-pair-flow}
v_t = (\R-1)P
\end{equation}
holding for an arbitrary differential function $P(v,v_x,v_{xx},\ldots)$,
where $\R$ is the mKdV recursion operator \eqref{mkdv-Rop}. 
When $P=-v_x$ is given by the $x$-translation symmetry 
$\X=P\partial/\partial_v$ of this recursion operator, 
the zero-curvature equations \eqref{real-akns-grad-eqn}--\eqref{real-akns-aux-eqn} yield
$K= -\tfrac{1}{2} v^2 + c$ and $\omega = -v_{xx} -\tfrac{1}{2}v^3 +c v$, 
where $c$ is an arbitrary constant given by the freedom in $D\inv_x$. 
The flow equation \eqref{mkdv-lax-pair-flow} then gives 
\begin{equation}
v_t = (c+1)v_x - \tfrac{3}{2}v^2 v_x -v_{xxx}
\end{equation}
which is the mKdV equation \eqref{mkdv-eqn} up to a convective term $(c+1)v_x$. 
This term vanishes if $c=-1$,
whereby $K= -\tfrac{1}{2} v^2 - 1$ and $\omega = -v_{xx} -\tfrac{1}{2}v^3 -v$,
so thus the matrices \eqref{real-akns-U}--\eqref{real-akns-V} become 
\begin{equation}
U = \tfrac{1}{2} 
\begin{pmatrix}
1 & v \\ -v & -1
\end{pmatrix}, 
\quad
V = \tfrac{1}{2} 
\begin{pmatrix}
-\tfrac{1}{2} v^2 - 1 & -v_{xx} -\tfrac{1}{2}v^3 -v -v_x
\\
v_{xx} +\tfrac{1}{2}v^3 +v -v_x & \tfrac{1}{2} v^2 + 1
\end{pmatrix} . 
\end{equation}
These matrices satisfy the zero-curvature equation \eqref{zero-curv} 
whenever $v$ is a solution of the mKdV equation \eqref{mkdv-eqn}. 
We can obtain a matrix Lax pair from them 
by applying a gauge transformation that consists of 
the mKdV scaling symmetry 
$v\to\lambda v$, $x\to\lambda^{-1}x$, $t\to\lambda^{-3}t$,
combined with the scaling 
$U\to\lambda U$, $V\to\lambda^3 V$. 
This is easily seen to produce the standard mKdV matrix Lax pair
\begin{equation}\label{mkdv-laxpair}
U = \tfrac{1}{2} 
\begin{pmatrix}
\lambda & v \\ -v & -\lambda
\end{pmatrix}, 
\quad
V = \tfrac{1}{2} 
\begin{pmatrix}
-\tfrac{1}{2}\lambda v^2 - \lambda^3 & -v_{xx} -\tfrac{1}{2}v^3 -\lambda^2 v -\lambda v_x
\\
v_{xx} +\tfrac{1}{2}v^3 +\lambda^2 v -\lambda v_x & \tfrac{1}{2}\lambda v^2 + \lambda^3
\end{pmatrix}
\end{equation}
up to a rescaling of the spectral parameter $\lambda$. 

We now use the relation \eqref{mkdv-recursion-rel} 
between the mKdV recursion operator 
and the recursion operator \eqref{mkdv-peakon-Rop} of the mCH equation
to re-write the flow equation \eqref{mkdv-lax-pair-flow} coming from the zero-curvature equation. 
This gives 
\begin{equation}\label{mkdv-lax-pair-peakon-flow}
v_t = (\wtil\R-1)\Delta P = (\wtil\R-1)\wtil P 
\end{equation}
with 
\begin{equation}
P = \Delta\inv\wtil P . 
\end{equation}
Since $\wtil\R(-v_x) = \wtil P^{(1)}$ produces the mCH flow \eqref{mch-flow}
when $D\inv_x(0)=0$, 
we consider $\wtil P = -v_x$
and, correspondingly, $P=-u_x$. 
Then the zero-curvature equations \eqref{real-akns-grad-eqn}--\eqref{real-akns-aux-eqn} give 
$K=-\tfrac{1}{2}(u^2-u_x^2)+ c$ 
and $\omega = -\tfrac{1}{2}(u^2-u_x^2)v -u_{xx}+ cv$, 
where $c=D\inv_x(0)$ is an arbitrary constant. 
Hence, the flow equation \eqref{mkdv-lax-pair-peakon-flow} becomes 
\begin{equation}
v_t = (\wtil\R-1)(-v_x) = (c+1)v_x -\tfrac{1}{2}((u^2-u_x^2)v)_x
\end{equation}
which is the mCH equation \eqref{mch-eqn} up to a convective term $(c+1)v_x$. 
We put $c=-1$ to make this term vanish, 
which yields 
\begin{equation}
K= -\tfrac{1}{2}(u^2-u_x^2)-1, 
\quad
\omega = -\tfrac{1}{2}(u^2-u_x^2)v -u . 
\end{equation}
The matrices \eqref{real-akns-U}--\eqref{real-akns-V} thereby 
satisfy the zero-curvature equation \eqref{zero-curv} 
whenever $v$ is a solution of the mCH equation \eqref{mch-eqn}. 
In particular, they are given by 
\begin{equation}
U = \tfrac{1}{2} 
\begin{pmatrix}
1 & v \\ -v & -1
\end{pmatrix}, 
\quad
V = \tfrac{1}{4} 
\begin{pmatrix}
u_x^2-u^2 -2 & (u_x^2-u^2)v -2(u +u_x)
\\
(u^2-u_x^2)v +2(u -u_x) & u^2-u_x^2 +2
\end{pmatrix} . 
\end{equation}
A Lax pair is obtained from these matrices 
by applying a gauge transformation defined by 
the mCH scaling symmetry 
$u\to\lambda^{-1} u$, $x\to x$, $t\to\lambda^{2}t$, 
combined with the scaling 
$U\to U$, $V\to\lambda^{-2} V$. 
This yields 
\begin{align}
U & = \tfrac{1}{2} 
\begin{pmatrix}
1 & \lambda v \\ -\lambda v & -1
\end{pmatrix},
\label{mch-U}
\\
V & = \tfrac{1}{4} 
\begin{pmatrix}
u_x^2 - u^2 -2\lambda^{-2} 
& \lambda(u_x^2 -u^2)v -2\lambda^{-1} (u +u_x)
\\
\lambda(u^2-u_x^2)v +2\lambda^{-1} (u -u_x) 
& u^2-u_x^2 + 2\lambda^{-2}
\end{pmatrix},
\label{mch-V}
\end{align}
which is the standard mCH matrix Lax pair 
(up to rescaling of the spectral parameter $\lambda$).

Thus, we have shown how the AKNS zero-curvature equation can be used to 
derive the matrix Lax pair for the mCH equation. 
This derivation has not appeared previously in the literature. 

An important final remark arises when we compare this Lax pair \eqref{mch-U}--\eqref{mch-V}
to the mKdV Lax pair \eqref{mkdv-laxpair}. 
In both of these Lax pairs, 
$U$ depends linearly on $\lambda$, 
while $V$ contains negative powers of $\lambda$ in the case of the mCH Lax pair
but only positive powers of $\lambda$ in the case of the mKdV Lax pair. 
This indicates that the mCH equation can be viewed as a negative flow in the mKdV hierarchy.

\section{Tri-Hamiltonian splitting in the NLS hierarchy}
\label{splitting}

The tri-Hamiltonian splitting method was originally applied to the NLS hierarchy 
in \Ref{OlvRos} by considering the standard bi-Hamiltonian structure of the NLS equation.
However, this did not lead to a peakon equation, 
because the operator $\Delta=1-D_x^2$ did not appear 
when the split Hamiltonian operators were recombined. 
We will show how to overcome this obstacle 
by using a third Hamiltonian structure of the NLS equation,
which is connected with the first higher flow in the NLS hierarchy. 

\subsection{NLS hierarchy}

We begin from the NLS equation 
\begin{equation}\label{nls-eqn}
\s\inum v_t + \tfrac{1}{2}|v|^2 v +v_{xx}=0
\end{equation}
which has the tri-Hamiltonian structure
\begin{equation}\label{nls-triHam-eqn}
v_t=\s\inum(\tfrac{1}{2}|v|^2 v +v_{xx})
= \I(\delta I/\delta\bar v)
= \H(\delta H/\delta\bar v)
=\D(\delta D/\delta\bar v)
\end{equation}
where
\begin{align}
\I & =\s \inum , 
\label{nls-Iop}
\\
\H & =-D_x -\inum vD\inv_x\Im\bar v , 
\label{nls-Hop}
\\
\D & =\s\inum( D_x^2 +vD\inv_x\Re\bar v D_x +\inum D_xv D\inv_x\Im\bar v)
\label{nls-Dop}
\end{align}
are mutually compatible Hamiltonian operators,
with $\Re$ and $\Im$ viewed as algebraic operators, 
and where
\begin{align}
D & =\int_{-\infty}^{\infty} |v|^2\,dx , 
\\
H & =\int_{-\infty}^{\infty} \Im(\bar v v_x)\,dx , 
\\
I & =\int_{-\infty}^{\infty} \tfrac{1}{4} |v|^4 - |v_x|^2 \,dx
\end{align}
are the corresponding Hamiltonians. 

The two lowest order Hamiltonian operators \eqref{nls-Iop}--\eqref{nls-Hop} 
yield the hereditary recursion operator
\begin{equation}\label{nls-Rop}
\R=\H\I\inv=\s\inum(D_x+v D\inv_x \Re\bar v) . 
\end{equation}
This operator and all three of the Hamiltonian operators 
are invariant under $x$-translations applied to $v$,
as well as phase rotations applied to $v$, 
corresponding to the invariance of the NLS equation under 
infinitesimal $x$-translations and infinitesimal phase rotations, 
which are represented by 
the symmetry operators $\X=-v_x\partial_v$ and $\X=\s\inum v\partial_v$. 
These operators yield the respective flows
\begin{equation}\label{nls-root-flows}
P^{(1)} = -v_x,
\quad
P^{(0)} = \s\inum v
\end{equation}
which are related by the NLS recursion operator \eqref{nls-Rop}:
\begin{equation}\label{nls-root-flow-rel}
P^{(1)} = \R P^{(0)} . 
\end{equation}
The NLS equation corresponds to a higher flow
\begin{equation}
v_t=-\R({-v_x})=-\R^2(\s\inum v)= \s\inum(\tfrac{1}{2}|v|^2 v +v_{xx})
\end{equation}
where the square of the recursion operator is given by 
\begin{equation}\label{nls-Rsqop}
\R^2=-\D\I\inv=-(D_x^2+D_xv D\inv_x \Re\bar v +\inum v D\inv_x \Im\bar v D_x) . 
\end{equation}
Note that the reduction of this operator \eqref{nls-Rsqop}
under the reality condition $\bar v=v$ 
is given by the negative of the mKdV recursion operator \eqref{mkdv-Rop}. 
Consequently, 
for comparison with the mKdV hierarchy, 
the most natural way to generate higher flows 
using the NLS recursion operator \eqref{nls-Rop} is by dividing the NLS hierarchy 
into even and odd flows
\begin{align}
P^{(2n)} &=(-\R^2)^n(\s\inum v),
\quad 
n=1,2,\ldots
\label{nls-even-flows}
\\
P^{(2n+1)} &=(-\R^2)^n({-v_x}),
\quad
n=1,2,\ldots
\label{nls-odd-flows}
\end{align}
which are related by 
\begin{equation}
P^{(2n+1)} = \R P^{(2n)} . 
\end{equation}
Then all of the odd flows \eqref{nls-odd-flows} will admit a real reduction that yields 
a corresponding flow in the mKdV hierarchy,
whereas all of the even flows \eqref{nls-odd-flows} do not possess a real reduction. 

Every flow in the NLS hierarchy \eqref{nls-even-flows}--\eqref{nls-odd-flows}
inherits a tri-Hamiltonian structure, 
due to Magri's theorem. 
For the even flows, 
this structure is given by 
\begin{equation}\label{nls-even-triHam}
P^{(2n)} = \I(\delta E^{(n)}/\delta\bar v)
= \H(\delta H^{(n-1)}/\delta\bar v)
=\D(\delta E^{(n-1)}/\delta\bar v),
\quad
n=1,2,\ldots
\end{equation}
where the gradients of the Hamiltonians are generated by 
\begin{align}
\delta E^{(n)}/\delta\bar v & = (-\R^*{}^2)^n(v) = Q^{(2n)}, 
\quad
n=0,1,2,\ldots
\\
\delta H^{(n)}/\delta\bar v & = -(-\R^*{}^2)^n(\s\inum v_x) = Q^{(2n+1)}= -\R^*Q^{(2n)},
\quad
n=0,1,2,\ldots
\end{align}
and where $\R^*$ is the adjoint recursion operator 
\begin{equation}\label{nls-adRop}
\R^*=\I\inv\H=\s(\inum D_x -vD\inv_x \Im\bar v) . 
\end{equation}
The odd flows have a similar but enlarged Hamiltonian structure
\begin{equation}\label{nls-odd-triHam}
P^{(2n+1)} = -\I(\delta H^{(n)}/\delta\bar v)
= \H(\delta E^{(n)}/\delta\bar v)
=-\D(\delta H^{(n-1)}/\delta\bar v)
= \E(\delta E^{(n-1)}/\delta\bar v), 
\quad
n=1,2,\ldots
\end{equation}
where
\begin{equation}\label{nls-Eop}
\begin{aligned}
\E=-\R^2\H & 
=-\big( D_x^3 +D_xvD\inv_x\Re\bar vD_x 
+ \inum D_x^2vD\inv_x\Im\bar v + \inum vD\inv_x\Im\bar v D_x^2
\\&\qquad\qquad
+\tfrac{1}{2}\inum v(|v|^2 D\inv_x\Im\bar v + D\inv_x |v|^2 \Im\bar v) \big)
\end{aligned}
\end{equation}
is a fourth Hamiltonian operator which is mutually compatible with $\I$, $\H$, $\D$. 
In particular, 
the first higher flow in this hierarchy \eqref{nls-odd-triHam} is 
an integrable $U(1)$-invariant version of the mKdV equation given by 
\begin{equation}\label{hirota-eqn}
v_t = \R(\s\inum(\tfrac{1}{2}|v|^2 v +v_{xx})) = -\R^2(-v_x) 
= -v_{xxx}-\tfrac{3}{2}|v|^2v_x
\end{equation}
which is the Hirota equation \cite{Hir1973}. 
It possesses the quad-Hamiltonian structure 
\begin{equation}\label{hirota-quadHamil}
v_t = -v_{xxx}-\tfrac{3}{2}|v|^2v_x
= -\I(\delta H^{(1)}/\delta\bar v) 
= \H(\delta E^{(1)}/\delta\bar v) 
= -\D(\delta H^{(0)}/\delta\bar v) 
= \E(\delta E^{(0)}/\delta\bar v)  . 
\end{equation}
Its recursion operator is given by the NLS squared recursion operator \eqref{nls-Rsqop}.

In the NLS hierarchy, 
the real reductions of $\H$ and $\E$ 
match the mKdV Hamiltonian operators \eqref{mkdv-Hop}--\eqref{mkdv-Eop}, 
while the Hamiltonians $E^{(n)}$, $n=0,1,2,\ldots$ 
match the mKdV Hamiltonians \eqref{H-dens}, \eqref{mkdv-H-expr}, 
multiplied by a factor of $2$, 
with the real reduction of the variational derivative 
$\delta/\delta\bar v = \tfrac{1}{2}(\delta/\delta\Re v +\inum \delta/\delta\Im v)$
having a compensating factor of $1/2$. 

The root flow in the NLS hierarchy of even flows \eqref{nls-even-triHam} 
has the Hamiltonian structure
\begin{equation}\label{nls-evenroot-flow}
P^{(0)} = \s\inum v 
=\I(\delta E^{(0)}/\delta\bar v)
\end{equation}
while the root flow in the NLS hierarchy of odd flows \eqref{nls-odd-triHam} 
has the bi-Hamiltonian structure
\begin{equation}\label{nls-oddroot-flow}
P^{(1)} = -v_x 
=-\I(\delta H^{(0)}/\delta\bar v) 
= \H(\delta E^{(0)}/\delta\bar v) . 
\end{equation}
Both of these root flows possess another Hamiltonian structure
if, similarly to the mKdV case, 
$D\inv_x(0)$ is redefined by the addition of some non-zero constant,
so that $D\inv_x(0)=c\neq 0$. 
Then the relations 
$\H(0)=-\inum v D\inv_x(0)$ and $\D(0)=\s(\inum v -v_x)D\inv_x(0)$ 
yield the Hamiltonian structures 
\begin{equation}\label{nls-rootflows-othstruct}
-P^{(0)} = \s c^{-1}\H(0),
\quad
P^{(1)} = c^{-1}(\H(0) \s \D(0))
\end{equation}
where the corresponding Hamiltonians are trivial. 

Each higher flow in the combined NLS hierarchy 
corresponds to a higher symmetry operator 
$\X^{(n)}=P^{(n)}\partial_v$, $n=1,2,\ldots$, 
all of which represent infinitesimal symmetries that are admitted by 
every equation $v_t= P^{(n)}$,  $n=1,2,\ldots$, in the combined hierarchy. 
The Hamiltonians $E^{(n)}$, $H^{(n)}$, $n=0,1,2,\ldots$, 
of the even and odd flows in the combined hierarchy 
correspond to conservation laws $\frac{d}{dt} H^{(n)}=\frac{d}{dt} E^{(n)}=0$, 
all of which hold for every equation $v_t= P^{(n)}$,  $n=1,2,\ldots$, 
in the combined hierarchy. 
All of the higher symmetries and higher Hamiltonian densities are local expressions
in terms of $v$ and its $x$-derivatives. 

Similarly to the mKdV case, 
the general scaling method \cite{Anc03} shown in the appendix 
can be applied to derive the simple scaling formulas
\begin{align}
& E^{(n)} =\int_{-\infty}^{\infty} e^{(n)}\,dx,
\quad
e^{(n)} = \tfrac{2}{2n+1}D\inv_x\Re\big(\bar vD_x(-\R^2)^nv\big), 
\quad
n=0,1,2,\ldots , 
\label{nls-H-expr}
\\
& H^{(n)}=\int_{-\infty}^{\infty}h^{(n)}\,dx,
\quad
h^{(n)} = \tfrac{1}{n+1}D\inv_x\Im\big(\s\bar vD_x(-\R^2)^nv_x\big),
\quad
n=0,1,2,\ldots .
\label{nls-E-expr}
\end{align}

\subsection{Hamiltonian triples}

Using the preceding preliminaries, 
we will now proceed with splitting the second and third NLS Hamiltonian operators 
\eqref{nls-Hop} and \eqref{nls-Dop}. 
Under scaling, 
the second Hamiltonian operator \eqref{nls-Hop} splits into two operators 
\begin{equation}\label{nls-splitHops}
\H_1=-D_x,
\quad
\H_2 =  -\inum vD\inv_x\Im\bar v
\end{equation}
which are a compatible Hamiltonian pair. 
Likewise, 
the third Hamiltonian operator \eqref{nls-Dop} splits into two operators 
\begin{equation}\label{nls-splitDops}
\D_1=\s\inum D_x^2 ,
\quad
\D_2 = \s\inum(D\inv_x\Re\bar v D_x +\inum D_xv D\inv_x\Im\bar v) 
\end{equation}
which are a compatible Hamiltonian pair. 
The first Hamiltonian operator \eqref{nls-Iop}, which obviously cannot be split, 
is compatible with each of the two pairs \eqref{nls-splitHops} and \eqref{nls-splitDops}. 
Hence, this yields two different Hamiltonian triples:
\begin{equation}\label{nls-triple1}
\I,
\quad
\H_1,
\quad
\H_2
\end{equation}
and 
\begin{equation}\label{nls-triple2}
\I,
\quad
\D_1,
\quad
\D_2 . 
\end{equation}

In \Ref{OlvRos}, 
the first triple \eqref{nls-triple1} was used to obtain the compatible Hamiltonian pair 
$\wtil\H = \I \pm \H_1=\s\inum \mp D_x$
and $\wtil\D = \H_2 =  -\inum vD\inv_x\Im\bar v$,
where $\wtil\H = (1 \pm \I D_x) \I$ factorizes to yield 
a symmetric operator $\Upsilon=1 \pm \I D_x$ 
which is the counterpart of the operator $\Delta=1-D_x^2$ in the mKdV case. 
The NLS root even flow \eqref{nls-evenroot-flow} can be expressed as 
a bi-Hamiltonian flow with respect to the Hamiltonian operators $\wtil\H$ and $\wtil\D$
through the introduction of a potential $u$ given by $v= \Upsilon u = u \pm \s\inum u_x$:
\begin{equation}
P^{(0)} = \s\inum v = \wtil\H(u) = -(\s c^{-1})\wtil\D(0)
\end{equation} 
where $u= \wtil\H(\delta\wtil H^{(0)}/\delta\bar v)$ 
holds for the Hamiltonian 
\begin{equation}\label{nls-peakon-H0}
\wtil H^{(0)} = \int_{-\infty}^{\infty} \Re(\bar v u)\, dx
= \int_{-\infty}^{\infty} (|u|^2 \pm \s\Im(\bar u_x u))\, dx . 
\end{equation} 
Then, from Magri's theorem, 
the hereditary recursion operator $\wtil\R = \wtil\D\wtil\H\inv$ 
produces the integrable equation
\begin{equation}\label{OlvRos-eqn}
v_t = \wtil\R(\s\inum v) = \pm (\s\inum \tfrac{1}{2}|u|^2 v) , 
\end{equation}
which has a bi-Hamiltonian structure 
$v_t = \wtil\D(\delta\wtil H^{(0)}/\delta\bar v) = \wtil\H(\delta\wtil H^{(1)}/\delta\bar v)$
coming from 
$\wtil\R(\s\inum v) = \wtil\D(u) = \wtil\H(\Upsilon\inv(\pm\tfrac{1}{2}|u|^2 v))$
with $u= \delta\wtil H^{(0)}/\delta\bar v$
and $\Upsilon\inv(\pm\tfrac{1}{2}|u|^2 v) = \delta\wtil H^{(1)}/\delta\bar v$. 
The first Hamiltonian $\wtil H^{(0)}$ is given by expression \eqref{nls-peakon-H0},
while the second Hamiltonian $\wtil H^{(1)}$ can be obtained by applying 
the variational relation 
\begin{equation}
\Upsilon\frac{\delta}{\delta\bar v} = \frac{\delta}{\delta\bar u}
\end{equation}
to get $\delta\wtil H^{(1)}/\delta\bar u = \pm\tfrac{1}{2}|u|^2 v$, 
which yields
\begin{equation}\label{nls-peakon-H1}
\wtil H^{(1)}= \int_{-\infty}^{\infty} \pm\tfrac{1}{2}|u|^2\Re(\bar u v)\, dx
= \int_{-\infty}^{\infty} \tfrac{1}{2}|u|^2 (|u|^2+ \s\Im(\bar u_x u))\, dx . 
\end{equation} 
However, as discussed in \Ref{OlvRos}, 
this bi-Hamiltonian equation \eqref{OlvRos-eqn} 
is \underbar{not} a peakon equation,
as it does not contain the operator $\Delta$. 
We add the remark that the NLS root odd flow \eqref{nls-oddroot-flow} 
also does not lead to a peakon equation. 
In particular, 
the first Hamiltonian operator $\wtil\H$ yields 
\begin{equation}
P^{(1)} = -v_x = \wtil\H(-\s\inum u_x) 
\end{equation} 
where $-\s\inum u_x= \delta\wtil E^{(0)}/\delta\bar v$ 
holds for the Hamiltonian 
\begin{equation}\label{nls-peakon-E0}
\wtil E^{(0)}= \int_{-\infty}^{\infty} \s\Im(\bar v u_x)\, dx
= \int_{-\infty}^{\infty} \s\Im(\bar u u_x)\, dx . 
\end{equation} 
However, another Hamiltonian structure cannot be found for this flow
by using the second Hamiltonian operator $\wtil\D$. 

We will instead make use of the other compatible Hamiltonian pair 
\begin{equation}\label{hirota-peakon-ops}
\wtil\H  = \I -\D_1 =\s\inum(1-D_x^2),
\quad
\wtil\D = \D_2 =  \s\inum(vD\inv_x\Re\bar v D_x +\inum D_xv D\inv_x\Im\bar v)
\end{equation}
where the first operator has the factorization 
\begin{equation}
\wtil\H  =\I\Delta = \Delta\I, 
\quad
\Delta=1-D_x^2 . 
\end{equation}
The corresponding hereditary recursion operator is given by 
\begin{equation}\label{hirota-peakon-Rop}
\wtil\R  =\wtil\D\wtil\H\inv 
= D_xv D\inv_x\Re\bar v \Delta\inv + \inum v D\inv_x\Im\bar v D_x\Delta\inv . 
\end{equation}
Both of the NLS root flows \eqref{nls-root-flows} have a Hamiltonian structure
with respect to the first Hamiltonian operator in the pair \eqref{hirota-peakon-ops}:
\begin{align}
& P^{(0)} = \s\inum v = \wtil\H(\delta\wtil H^{(0)}/\delta\bar v) , 
\label{hirota-rootflow-even}
\\
& P^{(1)} = -v_x = \wtil\H(\delta\wtil E^{(0)}/\delta\bar v) , 
\label{hirota-rootflow-odd}
\end{align}
with 
$\delta \wtil H^{(0)}/\delta\bar v = \Delta\inv v = u$
and 
$\delta \wtil E^{(0)}/\delta\bar v = -\s\inum\Delta\inv v_x = -\s\inum u_x$,
where the potential $u$ is now given by 
\begin{equation}\label{peakon-u}
v= \Delta u = u- u_{xx}
\end{equation}
and $v$ is referred to as the momentum variable. 
By using the variational relation \eqref{mkdv-varders}, 
we can formulate the previous Hamiltonian gradients as 
$\delta \wtil H^{(0)}/\delta\bar u =v$
and 
$\delta \wtil E^{(0)}/\delta\bar u = -\s\inum v_x$, 
from which we obtain the Hamiltonians 
\begin{align}
& \wtil H^{(0)} = \int_{-\infty}^{\infty} \Re(\bar u v)\, dx
= \int_{-\infty}^{\infty} (|u|^2 +|u_x|^2)\,dx , 
\\
& \wtil E^{(0)}= \int_{-\infty}^{\infty} \s\Im(\bar u v_x)\, dx
= \int_{-\infty}^{\infty} \s\Im(\bar u_x (u_{xx}-u))\,dx
\end{align} 
after integration by parts. 
The second Hamiltonian operator in the pair \eqref{hirota-peakon-ops} 
gives the Hamiltonian structure
\begin{equation}
\wtil\D(0) = \s (c_1\inum v + c_2v_x) . 
\end{equation}
if we redefine $D\inv_x(0)$ by the addition of an arbitrary non-zero constant, 
similarly to the mKdV case, so that $D\inv_x(0)=c\neq 0$.
Note that we have introduced two separate constants $c_1$ and $c_2$ 
due to the two separate $D\inv_x$ terms that occur in $\wtil\D$. 
In this sense, 
we can view each of the NLS root flows \eqref{hirota-rootflow-even}--\eqref{hirota-rootflow-odd}
as having a second Hamiltonian structure, 
$P^{(0)} = \s\inum v = c_1^{-1}\wtil\E(0)$ and $P^{(1)} = -v_x = -\s c_2^{-1}\wtil\E(0)$,
with a trivial Hamiltonian.

\section{$U(1)$-invariant peakon equations}
\label{derivation}

We will now explicitly show that the recursion operator \eqref{hirota-peakon-Rop} 
obtained from splitting the third Hamiltonian structure of the NLS equation 
generates a bi-Hamiltonian equation 
when it is applied to each of the root flows \eqref{hirota-rootflow-even}--\eqref{hirota-rootflow-odd} in the NLS hierarchy, 
with $u$ defined to be the peakon potential \eqref{peakon-u}. 
This will yield two $U(1)$-invariant peakon equations. 

From the even root flow \eqref{hirota-rootflow-even}, 
we obtain the first higher flow 
\begin{equation}\label{hirota-evenflow}
\wtil P^{(0,1)} = \wtil\R(\s\inum v) 
= \s\inum\big( \tfrac{1}{2}(|u|^2-|u_x^2|)v +\inum(\Im(\bar u u_x)v)_x \big) 
\end{equation}
where $\wtil\R=\wtil\D\wtil\H\inv $ is the recursion operator \eqref{hirota-peakon-Rop}, 
and where $\wtil\D$, $\wtil\H$ are the recombined Hamiltonian operators \eqref{hirota-peakon-ops}. 
The bi-Hamiltonian structure of this flow \eqref{hirota-evenflow} arises from 
\begin{equation} 
\wtil\R(\s\inum v) 
= \wtil\D(\Delta\inv v) 
= \wtil\H(\Delta\inv( \tfrac{1}{2}(|u|^2-|u_x^2|)v +\inum(\Im(\bar u u_x)v)_x ))
\end{equation}
in accordance with Magri's theorem, 
by expressing 
\begin{equation}\label{hirota-H0-gradient}
\Delta\inv v = \delta \wtil H^{(0)}/\delta\bar v
\end{equation}
and 
\begin{equation}\label{hirota-H1-gradient}
\Delta\inv( \tfrac{1}{2}(|u|^2-|u_x^2|)v +\inum(\Im(\bar u u_x)v)_x )
= \delta \wtil H^{(1)}/\delta\bar v
\end{equation}
with $v=\Delta u$. 
These two gradients have an alternative formulation
\begin{align}
\delta\wtil H^{(0)}/\delta\bar u & = v , 
\label{hirota-H0-u-gradient}
\\
\delta\wtil H^{(1)}/\delta\bar u & = \tfrac{1}{2}(|u|^2-|u_x^2|)v +\inum(\Im(\bar u u_x)v)_x
\label{hirota-H1-u-gradient} 
\end{align}
through the variational identity \eqref{mkdv-varders}. 
We remark that existence of the Hamiltonians $\wtil H^{(0)}$ and $\wtil H^{(1)}$ 
requires that the right-hand side of each gradient expression \eqref{hirota-H0-u-gradient}--\eqref{hirota-H1-u-gradient}
satisfies the Helmholtz conditions \cite{Olv-book}, as explained in the appendix. 
The first gradient \eqref{hirota-H0-u-gradient} yields 
the same Hamiltonian appearing in the Hamiltonian structure \eqref{hirota-rootflow-even} of the even root flow:
\begin{equation}\label{hirota-peakon-H0}
\wtil H^{(0)} = \int_{-\infty}^{\infty} \Re(\bar u v)\, dx
= \int_{-\infty}^{\infty} |u|^2 +|u_x|^2\,dx .
\end{equation}
For the second gradient \eqref{hirota-H1-u-gradient}, 
we can straightforwardly verify that the Helmholtz conditions hold
by a direct computation, 
with $v=\Delta u$. 
Then, since this gradient is given by a homogeneous expression under scaling of $u$,
we can obtain the Hamiltonian by using the scaling formula shown in the appendix. 
This yields 
\begin{equation}\label{hirota-peakon-H1}
\begin{aligned}
\wtil H^{(1)} & = \int_{-\infty}^{\infty} \tfrac{1}{2}\Re\big(\bar u (\tfrac{1}{2}(|u|^2-|u_x^2|)v +\inum(\Im(\bar u u_x)v)_x)\big)\,dx
\\
& = \int_{-\infty}^{\infty} \tfrac{1}{4}(|u|^4-|u_x|^4) + \tfrac{1}{2}\Re(\bar u_x^2 uv)\,dx
\end{aligned}
\end{equation}
modulo boundary terms. 
Hence, the flow \eqref{hirota-evenflow} has the bi-Hamiltonian structure
\begin{equation}
\wtil P^{(0,1)} 
= \wtil\D(\delta\wtil H^{(0)}/\delta\bar v) 
= \wtil\H(\delta\wtil H^{(1)}/\delta\bar v) . 
\end{equation}
The corresponding flow equation $v_t = \wtil P^{(0,1)}$ is given by 
\begin{equation}\label{hirota-even-peakon-eqn}
\s \inum v_t +\inum(\Im(\bar u u_x)v)_x +\tfrac{1}{2} (|u|^2-|u_x^2|)v =0
\end{equation}
which is a NLS-type peakon equation, 
with the bi-Hamiltonian formulation
\begin{equation}
\s v_t = \tfrac{1}{2} \inum(|u|^2-|u_x^2|)v -(\Im(\bar u u_x)v)_x 
= \wtil\D(\delta\wtil H^{(0)}/\delta\bar v) 
= \wtil\H(\delta\wtil H^{(1)}/\delta\bar v) . 
\end{equation}

Similarly,
from the odd root flow \eqref{hirota-rootflow-odd}, 
we obtain the first higher flow 
\begin{equation}\label{hirota-oddflow}
\wtil P^{(1,1)} = \wtil\R(-v_x) 
= \wtil\D(\s\inum u_x) = -\tfrac{1}{2}((|u|^2-|u_x^2|)v)_x -\inum\Im(\bar u u_x)v . 
\end{equation}
The bi-Hamiltonian structure of this flow \eqref{hirota-oddflow} arises from 
\begin{equation}\label{hirota-E0-E1-gradient}
\wtil\R(-v_x) 
= \wtil\D(\s\inum\Delta\inv v_x) 
= \wtil\H(\Delta\inv( \s\inum\tfrac{1}{2}((|u|^2-|u_x^2|)v)_x-\Im(\bar u u_x)v ))
\end{equation}
by expressing 
$\s\inum\Delta\inv v_x$ and $\s\inum\tfrac{1}{2}((|u|^2-|u_x^2|)v)_x-\Im(\bar u u_x)v$ 
in the gradient form 
\begin{align}
\delta\wtil E^{(0)}/\delta\bar u & = \s\inum v_x , 
\label{hirota-E0-u-gradient}
\\
\delta\wtil E^{(1)}/\delta\bar u & = \s\inum\tfrac{1}{2}((|u|^2 -|u_x^2|)v)_x -\Im(\bar u u_x)v
\label{hirota-E1-u-gradient}
\end{align}
through the variational identity \eqref{mkdv-varders}. 
As in the case of the previous flow, 
existence of the Hamiltonians $\wtil E^{(0)}$ and $\wtil E^{(1)}$ 
requires that the right-hand side of each gradient expression \eqref{hirota-E0-u-gradient}--\eqref{hirota-E1-u-gradient} 
satisfies the Helmholtz conditions. 
The first gradient \eqref{hirota-E0-u-gradient} yields 
the Hamiltonian that appears in the Hamiltonian structure \eqref{hirota-rootflow-odd} of the odd root flow:
\begin{equation}\label{hirota-peakon-E0}
\wtil E^{(0)}= \int_{-\infty}^{\infty} \s\Im(\bar u_x v)\, dx
= \int_{-\infty}^{\infty} \s\Im(\bar u_x (u_{xx}-u))\,dx . 
\end{equation}
For the second gradient \eqref{hirota-E1-u-gradient}, 
we can straightforwardly verify that the Helmholtz conditions hold
by a direct computation, with $v=\Delta u$. 
Then, since this gradient is given by a homogeneous expressions under scaling of $u$,
we can obtain the Hamiltonian by using the scaling formula shown in the appendix. 
This yields 
\begin{equation}\label{hirota-peakon-E1}
\begin{aligned}
\wtil E^{(1)} & = \int_{-\infty}^{\infty} \tfrac{1}{2}\Re\big(\bar u( \s\inum\tfrac{1}{2}((|u|^2 -|u_x^2|)v)_x -\Im(\bar u u_x)v )\big)\,dx
\\
& = -\int_{-\infty}^{\infty} \tfrac{3}{4}|u|^2\Im(\bar u u_x) +\tfrac{1}{4}(2|u|^2-|u_x|^2)\Im(\bar u_x u_{xx}) \,dx
\end{aligned}
\end{equation}
modulo boundary terms. 
Hence, the flow \eqref{hirota-oddflow} has the bi-Hamiltonian structure
\begin{equation}
\wtil P^{(1,1)} 
= \wtil\D(\delta\wtil E^{(0)}/\delta\bar v) 
= \wtil\H(\delta\wtil E^{(1)}/\delta\bar v) . 
\end{equation}
The corresponding flow equation $v_t = \wtil P^{(1,1)}$ is given by 
\begin{equation}\label{hirota-odd-peakon-eqn}
v_t +\tfrac{1}{2}((|u|^2-|u_x^2|)v)_x +\inum\Im(\bar u u_x)v
=0
\end{equation}
which is a Hirota-type peakon equation,
with the bi-Hamiltonian formulation
\begin{equation}
v_t= -\tfrac{1}{2}((|u|^2-|u_x^2|)v)_x -\inum\Im(\bar u u_x)v
= \wtil\D(\delta\wtil E^{(0)}/\delta\bar v) 
= \wtil\H(\delta\wtil E^{(1)}/\delta\bar v) . 
\end{equation}

Both of these peakon equations \eqref{hirota-even-peakon-eqn} and \eqref{hirota-odd-peakon-eqn}
are invariant under phase rotations 
\begin{equation}\label{phaserot-inv}
v\rightarrow e^{\s\inum\phi} v,
\quad
\phi=\const
\end{equation}
generated by $\X=\s\inum v\partial_v$,
where $u=\Delta\inv v$ transforms in the same way as $v$. 

Under the reality condition $\bar v=v$,
the first peakon equation \eqref{hirota-even-peakon-eqn} becomes trivial, 
while the second peakon equation \eqref{hirota-odd-peakon-eqn} 
reduces to the mCH peakon equation  \eqref{mch-eqn}. 

These two peakon equations \eqref{hirota-even-peakon-eqn} and \eqref{hirota-odd-peakon-eqn} 
can be formulated as $2$-component integrable systems by 
decomposing $u$ and $v$ into their real and imaginary parts:
\begin{equation}
u= u_1 + \inum u_2,
\quad
v= v_1 + \inum v_2
\end{equation}
with 
\begin{equation}
u_1 = \Re u,
\quad
u_2 = \Im u,
\quad
v_1 = \Re v,
\quad
v_2 = \Im v . 
\end{equation}
Then, 
the NLS-type peakon equation \eqref{hirota-even-peakon-eqn} 
is equivalent to the integrable system 
\begin{equation}
\s v_1{}_t +  Av_2 +(Bv_1)_x =0,
\quad
\s v_2{}_t  -Av_1 +(B v_2)_x =0
\end{equation}
and the Hirota-type peakon equation \eqref{hirota-odd-peakon-eqn} 
is equivalent to the integrable system 
\begin{equation}
v_1{}_t +  (Av_1)_x -Bv_2  =0,
\quad
v_2{}_t  +(Av_2)_x +B v_1  =0
\end{equation}
where
\begin{equation}
A = \tfrac{1}{2}( u_1^2+u_2^2 -(u_1{}_x)^2 -(u_2{}_x)^2 ),
\quad
B= u_1 u_2{}_x-u_2 u_1{}_x . 
\end{equation}
Each of these integrable systems is invariant under $SO(2)$ rotations on $(v_1,v_2)$,
with $(u_1,u_2)$ transforming the same way. 

Any linear combination of the $U(1)$-invariant bi-Hamiltonian peakon equations
\eqref{hirota-even-peakon-eqn} and \eqref{hirota-odd-peakon-eqn}
is again a $U(1)$-invariant bi-Hamiltonian peakon equation
\begin{equation}\label{hirota-combined-peakon-eqn}
\begin{aligned}
v_t & = -c_1( (Av)_x +\inum Bv ) + \s \inum c_2( Av+\inum (Bv)_x )
\\
& = \wtil\D(\delta(c_1\wtil E^{(0)}+c_2\wtil H^{(0)})/\delta\bar v) 
= \wtil\H(\delta(c_1\wtil E^{(1)}+c_2\wtil H^{(1)})/\delta\bar v) 
\end{aligned}
\end{equation}
with 
\begin{equation}
A = \tfrac{1}{2}(|u|^2 -|u_x|^2), 
\quad
B= \Im(\bar u u_x)
\end{equation}
where $c_1,c_2$ are arbitrary real constants. 
An equivalent form of this peakon equation \eqref{hirota-combined-peakon-eqn}
is given by 
\begin{equation}
\begin{aligned}
v_t & = \tilde c_1((A +\s B)v)_x + \tilde c_2 \inum(A+\s B)v
\\
& = -\wtil\D\big( \delta\big( \tfrac{1}{2}(\tilde c_1+\tilde c_2)\wtil E^{(0)} + \tfrac{1}{2}(\tilde c_1-\tilde c_2)\wtil H^{(0)} \big)/\delta\bar v\big)
\\
& = -\wtil\H\big( \delta\big( \tfrac{1}{2}(\tilde c_1+\tilde c_2)\wtil E^{(1)} + \tfrac{1}{2}(\tilde c_1-\tilde c_2)\wtil H^{(1)} \big)/\delta\bar v\big)
\end{aligned}
\end{equation}
where $\tilde c_1,\tilde c_2$ are arbitrary real constants. 

We will conclude this derivation by pointing out a simple relationship between 
the recursion operator \eqref{hirota-peakon-Rop} 
used in obtaining the $U(1)$-invariant peakon equations \eqref{hirota-even-peakon-eqn} and \eqref{hirota-odd-peakon-eqn},
and the NLS recursion operator \eqref{nls-Rop}. 
First, 
we express the split Hamiltonian operators \eqref{nls-splitDops} 
in terms of the recombined Hamiltonian operators \eqref{hirota-peakon-ops} 
by 
\begin{equation}
\D_1 = (1-\Delta) \I, 
\quad
\D_2 =\wtil\D . 
\end{equation} 
Next, we substitute these operator expressions 
into the NLS squared recursion operator \eqref{nls-Rsqop}, 
which is also the recursion operator of the Hirota equation \eqref{hirota-eqn}. 
This yields 
\begin{equation}\label{nls-Rrel1}
-\R^2=\D\I\inv=\D_1\I\inv +\D_2\I\inv
\end{equation}
where
\begin{equation}\label{nls-Rrel2}
\D_1\I\inv = 1-\Delta, 
\quad
\wtil\D_2\I\inv = \wtil\D\wtil\H\inv\Delta = \wtil\R\Delta . 
\end{equation}
Then, we combine these two equations \eqref{nls-Rrel1}--\eqref{nls-Rrel2} to get 
the relation
\begin{equation}\label{nls-recursion-rel}
-(\R^2+1)=(\wtil\R-1)\Delta . 
\end{equation}
Since $\wtil\R$ is a hereditary recursion operator, 
so is $\wtil\R-1$, and therefore 
\begin{equation}\label{hirota-nls-Rop}
\what\R = -(\R^2+1)\Delta\inv
\end{equation}
defines a hereditary recursion operator,
where the hereditary property \eqref{hereditaryprop} 
can be checked to hold directly, 
without the need for splitting the NLS Hamiltonian operators. 

The NLS-type peakon equation \eqref{hirota-even-peakon-eqn} 
arises from applying this recursion operator \eqref{hirota-nls-Rop}
to the even root flow \eqref{hirota-rootflow-even}, giving 
\begin{equation}
\hat P^{(0,1)} = \what\R(\s\inum v)  = -(\R^2+1)(\s \inum u)
\end{equation}
where 
\begin{equation}
\R^2(\s\inum u) = \s\inum( - u_{xx} - \tfrac{1}{2}(u^2-u_x^2)v +\inum (\Im(\bar u_x u)v)_x ) . 
\end{equation}
Hence, we have 
\begin{equation}\label{hirota-variant-evenflow}
\hat P^{(0,1)} = -(\R^2+1)(\s \inum u) 
= -\s\inum v+ \s\tfrac{1}{2} \inum(|u|^2-|u_x^2|)v - \s (\Im(\bar u u_x)v)_x 
= \wtil P^{(0,1)} - P^{(0)} 
\end{equation}
which is a linear combination of the root flow \eqref{hirota-rootflow-even}
and the NLS-type peakon flow \eqref{hirota-even-peakon-eqn}. 
Since both of those flows possess a bi-Hamiltonian structure 
given by the pair of Hamiltonian operators \eqref{hirota-peakon-ops}, 
the resulting flow \eqref{hirota-variant-evenflow} is also bi-Hamiltonian. 
The corresponding bi-Hamiltonian equation $v_t=\hat P^{(0,1)}$ 
is a slight generalization of the NLS-type peakon equation \eqref{hirota-even-peakon-eqn}:
\begin{equation}\label{hirota-variant-even-peakon-eqn}
\s \inum v_t -v +\inum (\Im(\bar u u_x)v)_x + \tfrac{1}{2} (|u|^2-|u_x^2|)v =0 . 
\end{equation}
In particular, 
a phase transformation $v\rightarrow \tilde v=e^{-\s\inum t}v$
applied to the NLS-type peakon equation \eqref{hirota-even-peakon-eqn}
yields the generalized equation \eqref{hirota-variant-even-peakon-eqn}. 

In a similar way, 
the Hirota-type peakon equation \eqref{hirota-odd-peakon-eqn} 
arises from applying the recursion operator \eqref{hirota-nls-Rop}
to the odd root flow \eqref{hirota-rootflow-odd}, giving  
\begin{equation}
\hat P^{(1,1)} = \what\R(-v_x)  = -(\R^2+1)(-u_x)
\end{equation}
where 
\begin{equation}
\R^2(-u_x) = u_{xxx} + \tfrac{1}{2}((u^2-u_x^2)v)_x +\inum \Im(\bar u_x u)v . 
\end{equation}
This yields a linear combination of the root flow \eqref{hirota-rootflow-odd}
and the Hirota-type peakon flow \eqref{hirota-odd-peakon-eqn},
\begin{equation}\label{hirota-variant-oddflow}
\hat P^{(1,1)} = -(\R^2+1)(-u_x) 
= v_x -\tfrac{1}{2} ((|u|^2-|u_x^2|)v)_x +\inum\Im(\bar u u_x)v
= \wtil P^{(1,1)} - P^{(1)} 
\end{equation}
which is a bi-Hamiltonian flow
with respect to the pair of Hamiltonian operators \eqref{hirota-peakon-ops}. 
The corresponding bi-Hamiltonian equation $v_t=\hat P^{(1,1)}$ 
is a slight generalization of the Hirota-type peakon equation \eqref{hirota-odd-peakon-eqn}:
\begin{equation}\label{hirota-variant-odd-peakon-eqn}
v_t -v_x  + \tfrac{1}{2} ((|u|^2-|u_x^2|)v)_x -\inum\Im(\bar u u_x)v =0 . 
\end{equation}
In particular, 
a Galilean transformation $x\rightarrow \tilde x=x+t$, $t\rightarrow\tilde t=t$
applied to the Hirota-type peakon equation \eqref{hirota-odd-peakon-eqn}
yields the generalized equation \eqref{hirota-variant-odd-peakon-eqn}. 

Finally, 
although the two root flows \eqref{hirota-rootflow-even} and \eqref{hirota-rootflow-odd} 
are related by the NLS recursion operator \eqref{nls-Rop}, 
the same relation does not hold for 
the two $U(1)$-invariant peakon flows \eqref{hirota-variant-evenflow} and \eqref{hirota-variant-oddflow}, 
since $-u_x - \R(\s\inum u) = -\inum\Im(\bar u_x u)v\neq0$
shows that $P^{(1,1)}\neq \R\hat P^{(0,1)}$.

\section{Integrability properties}
\label{laxpair}

We will now derive some further aspects of the integrability structure of 
the two $U(1)$-invariant peakon equations \eqref{hirota-even-peakon-eqn} and \eqref{hirota-odd-peakon-eqn}, 
specifically, 
their hierarchies of symmetries and conservation laws, 
and their Lax pairs. 

\subsection{Symmetries}

The recursion operator \eqref{hirota-peakon-Rop} 
which generates the $U(1)$-invariant peakon equations 
\eqref{hirota-even-peakon-eqn} and \eqref{hirota-odd-peakon-eqn} 
by applying it to the even and odd root flows \eqref{nls-root-flows}
in the NLS hierarchy 
is invariant under $x$-translations and $U(1)$-phase rotations. 
Consequently, 
from a standard result in the theory of recursion operators \cite{Olv77,Olv-book}, 
this operator $\wtil\R$ gives rise to two hierarchies of flows 
\begin{align}
\wtil P^{(0,n)} & = \wtil\R^n(\s\inum v),
\quad
n=1,2,\ldots , 
\label{peakon-even-flows}
\\
\wtil P^{(1,n)} & = \wtil\R^n(-v_x),
\quad
n=1,2,\ldots , 
\label{peakon-odd-flows}
\end{align}
starting from the two NLS root flows $P^{(0)}=\s\inum v$ and $P^{(1)}=-v_x$,
where the $n=1$ flows correspond to the respective $U(1)$-invariant peakon equations 
\eqref{hirota-even-peakon-eqn} and \eqref{hirota-odd-peakon-eqn}, 
as given by 
\begin{align}
v_t & = \wtil P^{(0,1)} 
= \s\tfrac{1}{2} \inum(|u|^2-|u_x^2|)v -\s(\Im(\bar u u_x)v)_x , 
\label{peakon-even-eqn}
\\
v_t & = \wtil P^{(1,1)} 
= -\tfrac{1}{2}((|u|^2-|u_x^2|)v)_x -\inum\Im(\bar u u_x)v . 
\label{peakon-odd-eqn}
\end{align}
Every flow in these two hierarchies is invariant under 
$x$-translations and $U(1)$-phase rotations. 
These two symmetries are respectively generated by 
\begin{equation}
\X_{\rm trans.} = -v_x\partial/\partial_v,
\quad
\X_{\rm phas.} = \inum v\partial/\partial_v . 
\end{equation}

In each hierarchy \eqref{peakon-even-flows} and \eqref{peakon-odd-flows}, 
every higher flow corresponds to a higher symmetry operator 
\begin{align}
\X^{(0,n)}& =\wtil P^{(0,n)}\partial/\partial_v,
\quad
n=2,3,\ldots
\label{peakon-even-symms}
\\
\X^{(1,n)}& =\wtil P^{(1,n)}\partial/\partial_v,
\quad
n=2,3,\ldots
\label{peakon-odd-symms}
\end{align}
representing an infinitesimal symmetry that is admitted by 
the respective $U(1)$-invariant peakon equations 
\eqref{peakon-even-eqn} and \eqref{peakon-odd-eqn}. 
All of these higher symmetries are nonlocal. 
In particular, 
the first higher symmetry in each hierarchy is given by 
\begin{equation}
\begin{aligned}
\wtil P^{(0,2)} & = 
\s\inum v\big( \Re(\bar u w-\bar u_x w_x) -\Im(\bar u W_x -\bar u_x W) + \tfrac{1}{2}(\Im(\bar u u_x))^2 -\tfrac{1}{4}(|u|^2-|u_x|^2) \big)
\\&\qquad
+\big(v\big( \Re(\bar u W-\bar u_x W_x) +\Im(\bar u w_x -\bar u_x w) - \Im(\bar u u_x)(|u|^2-|u_x|^2) \big)\big)_x
\end{aligned}
\end{equation}
and
\begin{equation}
\begin{aligned}
\wtil P^{(1,2)} & = 
-\big(v\big( \Re(\bar u w-\bar u_x w_x) -\Im(\bar u W_x -\bar u_x W) + \tfrac{1}{2}(\Im(\bar u u_x))^2 -\tfrac{1}{4}(|u|^2-|u_x|^2) \big)\big)_x
\\&\qquad
+ \inum v\big( \Re(\bar u W-\bar u_x W_x) +\Im(\bar u w_x -\bar u_x w) - \Im(\bar u u_x)(|u|^2-|u_x|^2) \big)
\end{aligned}
\end{equation}
where $w$ and $W$ are potentials defined in terms of $\Delta$ by the relations
\begin{equation}\label{hirota-1stord-potentials}
\Delta w = (|u|^2-|u_x|^2)v,
\quad
\Delta W = \Im(\bar u u_x)v . 
\end{equation}

\subsection{Conservation laws}

Since the recursion operator \eqref{hirota-peakon-Rop} is hereditary, 
Magri's theorem indicates that both 
hierarchies of flows \eqref{peakon-even-flows} and \eqref{peakon-odd-flows} 
have a bi-Hamiltonian structure 
\begin{align}
\wtil P^{(0,n)} & 
= \wtil\D(\delta\wtil H^{(n-1)}/\delta\bar v) 
= \wtil\H(\delta\wtil H^{(n)}/\delta\bar v),
\quad
n=1,2,\ldots, 
\label{peakon-even-flows-biHamil}
\\
\wtil P^{(1,n)} & 
= \wtil\D(\delta\wtil E^{(n-1)}/\delta\bar v) 
= \wtil\H(\delta\wtil E^{(n)}/\delta\bar v),
\quad
n=1,2,\ldots . 
\label{peakon-odd-flows-biHamil}
\end{align}
The Hamiltonian gradients are given by 
\begin{align}
\delta H^{(n)}/\delta\bar v & = \R^*{}^n(u) = Q^{(0,n)}, 
\quad
n=0,1,2,\ldots , 
\\
\delta E^{(n)}/\delta v & = \R^*{}^n(\s\inum u_x) = Q^{(1,n)}, 
\quad
n=0,1,2,\ldots , 
\end{align}
starting from the respective gradients \eqref{hirota-H0-gradient} and \eqref{hirota-E0-E1-gradient},
by using the adjoint recursion operator 
\begin{equation}\label{hirota-peakon-adRop}
\wtil\R^*  =\wtil\H\inv \wtil\D
= \Delta\inv(v D\inv_x\Re\bar v D_x + \inum D_x v D\inv_x\Im\bar v) . 
\end{equation}
These gradients also have the formulation 
\begin{align}
\delta H^{(n)}/\delta\bar u & = \Delta\wtil\R^*{}^n(u) = \I\inv \R^n \I (v) = \I\inv P^{(0,n)},
\quad
n=0,1,2,\ldots , 
\label{hirota-H-u-gradient}
\\
\delta E^{(n)}/\delta v & = \Delta\wtil\R^*{}^n(\s\inum u_x) = \I\inv \R^n\I(\s\inum v_x) = \I\inv P^{(1,n)}, 
\quad
n=0,1,2,\ldots , 
\label{hirota-E-u-gradient}
\end{align}
which is obtained through the variational identity \eqref{mkdv-varders}
combined with the recursion operator relation 
\begin{equation}
\I\Delta\wtil\R^* = \wtil\R\I\Delta . 
\end{equation}
We can get a formula for the Hamiltonians by using the general scaling method \cite{Anc03}
shown in the appendix, 
which relies on the gradients \eqref{hirota-H-u-gradient} and \eqref{hirota-E-u-gradient}
being scaling-homogeneous expressions of $u$ and $x$-derivatives of $u$. 
This yields 
\begin{align}
\wtil H^{(n)} & = \int_{-\infty}^{\infty} \tilde h^{(n)}\,dx,
\quad
\tilde h^{(n)}= \tfrac{1}{n+1} \Im(\bar u P^{(0,n)}) , 
\label{hirota-peakon-H-Q-expr}
\\
\wtil E^{(n)} & = \int_{-\infty}^{\infty} \tilde e^{(n)}\,dx,
\quad
\tilde e^{(n)}= \tfrac{1}{n+1} \Im(\bar u P^{(1,n)})
\label{hirota-peakon-E-Q-expr}
\end{align}
modulo boundary terms. 
All of the higher Hamiltonian densities are nonlocal. 
In particular, 
the first higher density in each hierarchy is given by 
\begin{equation}
\begin{aligned}
\tilde h^{(2)} & = -\tfrac{1}{3}\s \Big( 
\Re(\bar u v)\big( \Re(\bar u w-\bar u_x w_x) -\Im(\bar u W_x -\bar u_x W) + \tfrac{1}{2}(\Im(\bar u u_x))^2 -\tfrac{1}{4}(|u|^2-|u_x|^2) \big)
\\&\qquad
+\Im(\bar u_x v)\big( \Re(\bar u W-\bar u_x W_x) +\Im(\bar u w_x -\bar u_x w) - \Im(\bar u u_x)(|u|^2-|u_x|^2) \big)
\Big)
\end{aligned}
\end{equation}
and
\begin{equation}
\begin{aligned}
\tilde e^{(2)} & = \tfrac{1}{3}\s \Big( 
\Im(\bar u_x v)\big( \Re(\bar u w-\bar u_x w_x) -\Im(\bar u W_x -\bar u_x W) + \tfrac{1}{2}(\Im(\bar u u_x))^2 -\tfrac{1}{4}(|u|^2-|u_x|^2) \big)
\\&\qquad
-\Re(\bar u v)\big( \Re(\bar u W-\bar u_x W_x) +\Im(\bar u w_x -\bar u_x w) - \Im(\bar u u_x)(|u|^2-|u_x|^2) \big)
\Big)
\end{aligned}
\end{equation}
where $w$ and $W$ are the potentials \eqref{hirota-1stord-potentials}. 

\subsection{Lax pair}

The recursion operator \eqref{hirota-peakon-Rop} generating 
the two hierarchies of $U(1)$-invariant peakon flows
\eqref{peakon-even-flows}--\eqref{peakon-odd-flows}
can be derived similarly to the mCH recursion operator 
by using a matrix zero-curvature equation \eqref{zero-curv}
based on the AKNS scheme. 
Here the zero-curvature matrices are taken to have the form 
\begin{align}
U & = 
\begin{pmatrix}
\lambda & \tfrac{1}{2} v
\\
-\tfrac{1}{2}\bar v & -\lambda
\end{pmatrix}, 
\label{akns-U}
\\
V & = 
\begin{pmatrix}
\lambda K +\tfrac{1}{2}\inum J & \tfrac{1}{2} \omega + \lambda P
\\
-\tfrac{1}{2}\bar \omega + \lambda\bar P & -\lambda K -\tfrac{1}{2}\inum J
\end{pmatrix} ,
\label{akns-V}
\end{align}
belonging to the Lie algebra $sl(2,\Cnum)$,
where $K$ and $J$ are real functions, 
and where $P$ and $\omega$ are complex functions. 
Note that $U$ is the standard AKNS matrix, 
but with the spectral parameter $\lambda$ being real, 
and $V$ is parameterized such that 
$\begin{pmatrix} K & 0 \\ 0 & -K \end{pmatrix}$,
$\begin{pmatrix} \inum J & 0 \\ 0 & -\inum J \end{pmatrix}$,
$\begin{pmatrix} 0 & \omega \\ -\bar\omega & 0 \end{pmatrix}$, 
$\begin{pmatrix} 0 & P \\ \bar P & 0 \end{pmatrix}$
are mutually orthogonal in the Cartan-Killing inner product in $sl(2,\Cnum)$.
(Recall, this inner product is defined by the trace of the product of 
a $sl(2,\Cnum)$ matrix and a hermitian conjugated $sl(2,\Cnum)$ matrix.)
In this representation, 
the components of the zero-curvature equation \eqref{zero-curv} yield
\begin{subequations}
\begin{align}
v_t & = D_x \omega + \inum J v  -4\lambda^2 P, 
\label{akns-flow-eqn}
\\
\omega & = K v + D_x P , 
\label{akns-grad-eqn}
\\
D_x K & = \Re(\bar v P) , 
\label{akns-aux-eqn1}
\\
D_x J & = \Im(\bar v\omega) . 
\label{akns-aux-eqn2}
\end{align}
\end{subequations}
Equations \eqref{akns-grad-eqn}--\eqref{akns-aux-eqn1}
can be used to express 
$K=D\inv_x\Re(\bar v P)$ and $\omega = v D\inv_x\Re(\bar v P) + D_x P$,
and equation \eqref{akns-aux-eqn2} then gives
$J = D\inv_x\Im(\bar v\omega) = D\inv_x\Im(\bar v D_x P)$,
which yields 
$v_t = D_x^2P +D_x(vD\inv_x\Re(\bar vP)) + \inum v D\inv_x\Im(\bar v D_x P)  -4\lambda^2 P$
from equation \eqref{akns-flow-eqn}. 
If we now put $\lambda =\tfrac{1}{2}$ then we obtain 
\begin{equation}\label{hirota-lax-pair-flow}
v_t = -(\R^2+1)P
\end{equation}
holding for an arbitrary differential function $P(v,v_x,v_{xx},\ldots)$,
where $\R$ is the NLS recursion operator \eqref{nls-Rop}
and $-\R^2$ is the recursion operator \eqref{nls-Rsqop} that generates 
the even flows \eqref{nls-even-flows} and the odd flows \eqref{nls-odd-flows} 
in the NLS hierarchy. 
This shows that, in the matrices \eqref{akns-U}--\eqref{akns-V}, 
$P$ can be identified with the generator of any flow in the NLS hierarchy. 

As a first step to obtain a Lax pair in a systematic way, 
we will show how to use these matrices \eqref{akns-U}--\eqref{akns-V} 
to derive a Lax pair for the root flows 
in each of the two hierarchies \eqref{nls-even-flows} and \eqref{nls-odd-flows}.

In particular, 
for $P=\inum v$, which corresponds to the even root flow in the NLS hierarchy, 
the zero-curvature equations \eqref{akns-grad-eqn}--\eqref{akns-aux-eqn2} yield
$K= c_1$, $\omega = \inum v_x  +c_1 v$, $J= \tfrac{1}{2}|v|^2 +c_2$, 
where $c_1,c_2$ are arbitrary constants given by the freedom in $D\inv_x$. 
The flow equation \eqref{hirota-lax-pair-flow} then produces the NLS equation \eqref{nls-eqn}
if we choose $c_1=0$ and $c_2=-1$. 
Consequently, substitution of 
\begin{equation}
K=0,
\quad
\omega = \inum v_x ,
\quad
J= \tfrac{1}{2}|v|^2 -1
\end{equation}
into these matrices \eqref{akns-U}--\eqref{akns-V}, 
followed by a gauge transformation consisting of 
the NLS scaling symmetry 
$v\to\lambda v$, $x\to\lambda^{-1}x$, $t\to\lambda^{-2}t$
combined with the scaling 
$U\to\lambda U$, $V\to\lambda^2 V$, 
can be seen to give the standard NLS matrix Lax pair
\begin{equation}\label{nls-laxpair}
U = \tfrac{1}{2} 
\begin{pmatrix}
\lambda & v \\ -\bar v & -\lambda
\end{pmatrix}, 
\quad
V = \tfrac{1}{2} \inum
\begin{pmatrix}
\tfrac{1}{2}|v|^2 +\lambda^2 & v_x +\lambda v
\\
\bar v_x -\lambda \bar v &  - \tfrac{1}{2}|v|^2 -\lambda^2
\end{pmatrix}
\end{equation}
up to rescaling the spectral parameter. 
Likewise, for $P=-v_x$, which corresponds to the odd root flow in the NLS hierarchy, 
the zero-curvature equations \eqref{akns-grad-eqn}--\eqref{akns-aux-eqn2} yield
$K= -\tfrac{1}{2} |v|^2 + c_1$, $\omega = -v_{xx} -\tfrac{1}{2}|v|^2v +c_1 v$, 
$J=\Im(\bar v_x v) +c_2$. 
If we now choose $c_1=-1$ and $c_2=0$, 
then the flow equation \eqref{hirota-lax-pair-flow} produces the Hirota equation \eqref{hirota-eqn}. 
Substitution of 
\begin{equation}
K= -\tfrac{1}{2} |v|^2 - 1,
\quad
\omega = -v_{xx} -\tfrac{1}{2}|v|^2v -v,
\quad
J = \Im(\bar v_x v)
\end{equation}
into the matrices \eqref{akns-U}--\eqref{akns-V},
followed by a gauge transformation 
$v\to\lambda v$, $x\to\lambda^{-1}x$, $t\to\lambda^{-3}t$, 
$U\to\lambda U$, $V\to\lambda^3 V$, 
thereby gives a matrix Lax pair for the Hirota equation
\begin{equation}\label{hirota-laxpair}
U = \tfrac{1}{2} 
\begin{pmatrix}
\lambda & v \\ -\bar v & -\lambda
\end{pmatrix}, 
\quad
V = \tfrac{1}{2} 
\begin{pmatrix}
\inum\Im(\bar v_x v) -\tfrac{1}{2}\lambda |v|^2 - \lambda^3 
& -v_{xx} -\tfrac{1}{2}|v|^2v -\lambda^2 v -\lambda v_x
\\
\bar v_{xx} +\tfrac{1}{2}|v|^2\bar v +\lambda^2\bar v -\lambda\bar v_x 
& -\inum\Im(\bar v_x v) +\tfrac{1}{2}\lambda |v|^2  +\lambda^3 
\end{pmatrix}
\end{equation}
up to rescaling the spectral parameter. 

We will next adapt the previous steps to both 
the NLS-type peakon equation \eqref{hirota-even-peakon-eqn} 
and the Hirota-type peakon equation \eqref{hirota-odd-peakon-eqn}. 
The main idea is to use the NLS recursion operator relation \eqref{nls-recursion-rel} 
to express the zero-curvature flow equation \eqref{hirota-lax-pair-flow} 
in terms of the recursion operator $\wtil\R$ 
that generates the two hierarchies of $U(1)$-invariant peakon flows
\eqref{peakon-even-flows}--\eqref{peakon-odd-flows}. 
This yields 
\begin{equation}\label{hirota-lax-pair-peakon-flow}
v_t = (\wtil\R-1)\Delta P = (\wtil\R-1)\wtil P 
\end{equation}
with 
\begin{equation}
P = \Delta\inv\wtil P . 
\end{equation}
As a result, $P$ can be chosen to be the generator of any flow
in the hierarchies \eqref{peakon-even-flows}--\eqref{peakon-odd-flows}. 
We proceed by separately considering the two flows 
\begin{equation}
\wtil P^{(0,0)} = \inum v,
\quad
\wtil P^{(1,0)} = -v_x
\end{equation}
which are the respective ($n=0$) root flows in these two hierarchies. 
The corresponding flows on the potential $u$ are given by 
\begin{equation}
P^{(0,0)}= \inum u,
\quad
P^{(1,0)}=-u_x  . 
\end{equation}

For the first flow $P=P^{(0,0)} = \inum u$, 
the zero-curvature equations \eqref{akns-grad-eqn}--\eqref{akns-aux-eqn2} give 
$K=\Im(\bar u_x u)+ c_1$, 
$\omega = \Im(\bar u_x u)v+ c_1v +\inum u_x$, 
$J= \tfrac{1}{2}(|u|^2-|u_x|^2)+ c_2$. 
Hence, the flow equation \eqref{hirota-lax-pair-peakon-flow} becomes 
\begin{equation}
v_t = (\wtil\R-1)(\inum v) = \inum(c_2-1)v +c_1v_x +\tfrac{1}{2}\inum (|u|^2-|u_x|^2)v + (\Im(\bar u_x u)v)_x
\end{equation}
which is the NLS-type peakon equation \eqref{hirota-even-peakon-eqn} 
if we put $c_2=1$ and $c_1=0$. 
Then, by substituting 
\begin{equation}
K=\Im(\bar u_x u), 
\quad
\omega = \Im(\bar u_x u)v +\inum u_x,
\quad
J= \tfrac{1}{2}(|u|^2-|u_x|^2)+ 1
\end{equation}
into the matrices \eqref{akns-U}--\eqref{akns-V},
and applying a gauge transformation given by 
\begin{equation}\label{gaugetrans}
u\to\lambda^{-1} u,
\quad
x\to x,
\quad
t\to\lambda^2 t,
\quad
U\to U,
\quad
V\to\lambda^{-2} V, 
\end{equation}
we obtain the Lax pair 
\begin{align}
U & = \tfrac{1}{2} 
\begin{pmatrix}
1 & \lambda v \\ -\lambda\bar v & -1
\end{pmatrix},
\label{nls-peakon-U}
\\
V & = \tfrac{1}{4}
\begin{pmatrix}
\inum(|u|^2 -|u_x|^2) +2\Im(\bar u_x u) +2\inum\lambda^{-2}\ 
& 2\lambda\Im(\bar u_x u)v  +2\inum\lambda^{-1}(u +u_x)
\\
-2\lambda\Im(\bar u_x u) \bar v -2\inum\lambda^{-1}(\bar u -\bar u_x)\ 
& \inum(|u_x|^2 -|u|^2) -2\Im(\bar u_x u) -2\inum\lambda^{-2}
\end{pmatrix} . 
\label{nls-peakon-V}
\end{align}

For the second flow $P= P^{(1,0)} = -u_x$, 
the zero-curvature equations \eqref{akns-grad-eqn}--\eqref{akns-aux-eqn2} give 
$K=-\tfrac{1}{2}(|u|^2-|u_x|^2)+ c_1$, 
$\omega = -\tfrac{1}{2}(|u|^2-|u_x|^2)v+ c_1v -u_{xx}$, 
$J= \Im(\bar u_x u)+ c_2$. 
Hence, the flow equation \eqref{hirota-lax-pair-peakon-flow} becomes 
\begin{equation}
v_t = (\wtil\R-1)(\inum v) = \inum c_2v +(c_1+1)v_x -\tfrac{1}{2}((|u|^2-|u_x|^2)v)_x +\inum \Im(\bar u_x u)v
\end{equation}
which is the Hirota-type peakon equation \eqref{hirota-odd-peakon-eqn} 
if we put $c_1=-1$ and $c_2=0$. 
Then, by substituting 
\begin{equation}
K=-\tfrac{1}{2}(|u|^2-|u_x|^2) -1,
\quad
\omega = -\tfrac{1}{2}(|u|^2-|u_x|^2)v - u, 
\quad
J= \Im(\bar u_x u)
\end{equation}
into the matrices \eqref{akns-U}--\eqref{akns-V},
and applying the gauge transformation \eqref{gaugetrans}, 
we obtain the Lax pair 
\begin{align}
U & = \tfrac{1}{2} 
\begin{pmatrix}
1 & \lambda v \\ -\lambda\bar v & -1
\end{pmatrix} ,
\label{hirota-peakon-U}
\\
V & = \tfrac{1}{4} 
\begin{pmatrix}
|u_x|^2 -|u|^2 +2\inum \Im(\bar u_x u) -2\lambda^{-2}\ 
& \lambda(|u_x|^2-|u|^2)v  -2\lambda^{-1}(u +u_x)
\\
\lambda(|u|^2 -|u_x|^2)\bar v  +2\lambda^{-1}(\bar u -\bar u_x)
& |u|^2 -|u_x|^2 -2\inum \Im(\bar u_x u) +2\lambda^{-2}\ 
\end{pmatrix} . 
\label{hirota-peakon-V}
\end{align}

Comparing the NLS Lax pair \eqref{nls-laxpair} 
to these two Lax pairs \eqref{nls-peakon-U}--\eqref{nls-peakon-V} and \eqref{hirota-peakon-U}--\eqref{hirota-peakon-V}, 
we see that $U$ depends linearly on $\lambda$ 
while $V$ contains negative powers of $\lambda$ in the case of the peakon equations
but only positive powers of $\lambda$ in the case of the NLS equation. 
This indicates that the two peakon equations can be viewed as negative flows in the NLS hierarchy.

\section{Peakon solutions}
\label{peakons}

A peakon $u(t,x)$ is a peaked travelling wave
\begin{equation}\label{peakon}
u = a \exp(-|x-c t|),
\quad
a,c=\const
\end{equation}
where the amplitude $a$ and the speed $c$ are related by some algebraic equation. 
This expression \eqref{peakon} is motivated by the form of the kernel of the operator 
$\Delta = 1- \partial_x^2$. 
The corresponding momentum variable \eqref{peakon-u} is a distribution 
\begin{equation}
v = u-u_{xx} = 2a \delta(x-ct) . 
\end{equation}

Peakons are not classical solutions 
and instead are commonly understood as weak solutions
\cite{ConStr,ConMol2000,ConMol2001,GuiLiuTian,Hak,GuiLiuOlvQu}
in the setting of an integral (weak) formulation of a given peakon equation. 
A weak formulation is defined by multiplying the peakon equation by test function $\psi(t,x)$ 
and integrating by parts to remove all terms involving $v$ and derivatives of $v$,
leaving at most $u$, $u_x$, and $u_t$ in the integral. 
The weak formulation of the mCH equation \eqref{mch-eqn} is given by 
\begin{equation}\label{weak-CH}
0=\iint_{\Rnum^2} \Big(
(\psi-\psi_{xx})u_t + (3\psi -\psi_{xx})u u_x 
-\tfrac{1}{2} \psi_x u_x^2 
\Big) dx\; dt . 
\end{equation}
Its well-known peakon solution \cite{Qia06,GuiLiuOlvQu}
has the amplitude-speed relation $a=c$:
\begin{equation}\label{mCH-1peakon}
u = c \exp(-|x-c t|),
\quad
c=\const 
\end{equation}

For both the NLS-type peakon equation \eqref{peakon-even-eqn} 
and the Hirota-type peakon equation \eqref{peakon-odd-eqn},
their $U(1)$-invariance allows for the possibility of oscillatory peakon solutions
whose form is given by a peaked travelling wave $a\exp(-|x-c t|)$ modulated by 
an oscillatory plane-wave phase $\exp(\inum (\phi+\w t+\k x))$. 

We will begingby considering smooth plane-wave solutions
\begin{equation}\label{planewave}
u = a\exp(\inum(\k x+\w t)),
\quad
a,\k,\w=\const 
\end{equation}
Substitution of this expression into the Hirota-type peakon equation \eqref{peakon-odd-eqn} yields 
\begin{equation}
\w= \tfrac{1}{2}a^2 \k(\k^2-3)
\end{equation}
which represents a nonlinear dispersion relation for the plane-wave. 
Similarly, 
the NLS-type peakon equation \eqref{peakon-even-eqn} yields
\begin{equation}
\w= \tfrac{1}{2}a^2 (1-3\k^2)
\end{equation}
which is a different nonlinear dispersion relation. 
Note the speeds $c=-\w/\k$ of the resulting plane-waves are respectively given by 
\begin{equation}\label{hirota-planewave-rel}
c= \tfrac{1}{2}a^2 (3-\k^2)
\end{equation}
and
\begin{equation}\label{nls-planewave-rel}
c= \tfrac{1}{2}a^2 (3\k^2-1)/\k . 
\end{equation}

In the Hirota case \eqref{hirota-planewave-rel},
the plane-wave has the amplitude-speed form 
\begin{equation}\label{hirota-planewave}
u = a\exp\big( \pm\inum \sqrt{3-2c/a^2}(x-ct) \big)
\end{equation}
where $c \leq \tfrac{3}{2}a^2$. 
This wave can move in either direction 
but has a maximum speed of $c_{\max}= \tfrac{3}{2}a^2$ in the positive $x$ direction. 
In the NLS case \eqref{hirota-planewave-rel},
the amplitude-speed form of the plane-wave is given by 
\begin{equation}\label{nls-planewave}
u = a\exp\big( \inum \tfrac{1}{3}(\pm\sqrt{3+c^2/a^4}+c/a^2)(x-ct) \big) . 
\end{equation}
This wave can move in either direction, with no restriction on its speed. 

These features of the plane-wave solutions are analogous to the features of 
the oscillatory solitons \cite{AncMiaWil} of the Hirota equation \eqref{hirota-eqn} 
and the usual solitons \cite{AblKauNewSeg} of the NLS equation \eqref{nls-eqn}.

We will now seek oscillatory peakon solutions 
\begin{equation}\label{oscil-peakon}
u = a \exp(\inum (\phi+\w t+\k x))\exp(-|\xi|),
\quad
\xi= x-c t,
\quad
a,c,\phi,\w,\k=\const
\end{equation}
where $\w/(2\pi)$ is a temporal oscillation frequency, 
$2\pi/k$ is a spatial modulation length,
and $\phi$ is a phase angle. 
The momentum variable \eqref{peakon-u} 
corresponding to expression \eqref{oscil-peakon} is given by the distribution 
\begin{equation}
v = u-u_{xx} 
= 2 a \exp(\inum (\phi+\w t)) \delta(\xi) + a \exp(\inum (\phi+\w t+\k x) -|\xi|) \big( \inum 2\k\, \sgn(\xi) +\k^2 \big) .
\end{equation}

To proceed, we first observe that neither of the $U(1)$-invariant peakon equations \eqref{peakon-even-eqn} and \eqref{peakon-odd-eqn} 
has a weak formulation that involves only $u$, $\bar u$, and their first derivatives. 
In particular, after multiplication by a complex-valued test function $\psi(t,x)$, 
the NLS-type equation \eqref{peakon-even-eqn} contains problematic terms 
$\psi |u_x|^2u_{xx} = \tfrac{1}{2}\psi \bar u_x (u_x^2)_x$ 
and $\psi (\bar u_x uu_{xx})_x \equiv -\psi_x \bar u_x uu_{xx}$, 
which cannot be expressed as total $x$-derivatives
(where ``$\equiv$'' denotes equality modulo a total $x$-derivative). 
Likewise the Hirota-type equation \eqref{peakon-odd-eqn} contains problematic terms 
$\psi (|u_x|^2u_{xx})_x \equiv -\psi_x \bar u_x u_x u_{xx}$ 
and $\psi \bar u_x uu_{xx}$. 

But since the oscillatory peakon expression \eqref{oscil-peakon} 
factorizes into a standard peakon amplitude expression $a \exp(-|x-c t|)$ 
and an oscillatory phase expression $\exp(i(\phi+\w t+\k \xi))$, 
we can consider a weak formulation in which a polar form 
\begin{equation}\label{polarform}
u=A\exp(\inum\Phi),
\quad
\bar u=A\exp(-\inum\Phi)
\end{equation}
is used such that any derivatives of $A$ of second and higher orders are removed. 
Specifically, 
first we substitute this polar form into the peakon equations \eqref{peakon-even-eqn} and \eqref{peakon-odd-eqn} multiplied by the test function $\psi(t,x)$;
next we use integration by parts to remove all terms involving 
$A_{xx}$, $A_{tx}$, and their derivatives;
then we split the resulting integral equation into its real and imaginary parts. 
This yields what we will call a \emph{weak-amplitude} formulation. 

\begin{lem}\label{lem:weakform}
(i) 
The NLS-type peakon equation \eqref{peakon-even-eqn} in polar form 
has the weak-amplitude formulation
\begin{align}
& \begin{aligned}
0=&\iint_{\Rnum^2} \Big( 
\psi_{1xx} \big( A_t+\Phi_x A^2 A_x \big)
+\psi_{1x} \big( 2 \Phi_x AA_x^2+\Phi_{xx}A^2A_x \big)
-\psi_{1} \big( A_t +\tfrac{5}{6}\Phi_x A_x^3 
\\&\qquad
+\tfrac{1}{2} \Phi_{xx} AA_x^2 
+\tfrac{1}{2}(5+7\Phi_x^2)\Phi_xA^2 A_x 
+\tfrac{1}{2}(1+7\Phi_x^2)\Phi_{xx} A^3 \big)
\Big) dx\; dt, 
\end{aligned}
\label{weak-nls-peakon-eqn-Re}
\\
& \begin{aligned}
0=&\iint_{\Rnum^2} \Big( 
\psi_{2xx} A \Phi_t
-\psi_{2x}\big( \tfrac{1}{6}A_x^3 +\tfrac{1}{2}(3\Phi_x^2 -1)A^2A_x +\Phi_x\Phi_{xx}A^3 \big)
-\psi_{2}\big( \tfrac{1}{2}(3\Phi_x^2 -1)AA_x^2 
\\&\qquad
+\Phi_x\Phi_{xx}A^2A_x -\tfrac{1}{2}(1+\Phi_x^2)(1-3\Phi_x^2)A^3
+\Phi_t A \big)
\Big) dx\; dt, 
\end{aligned}
\label{weak-nls-peakon-eqn-Im}
\end{align}
where $\psi_{1}(t,x),\psi_{2}(t,x)$ are real test functions. 
This pair of integral equations \eqref{weak-nls-peakon-eqn-Re}--\eqref{weak-nls-peakon-eqn-Im} 
is satisfied by all classical solutions of the NLS-type peakon equation \eqref{peakon-even-eqn}. 
\newline
(ii) 
The Hirota-type equation \eqref{peakon-odd-eqn} in polar form 
has the weak-amplitude formulation
\begin{align}
& \begin{aligned}
0=&\iint_{\Rnum^2} \Big( 
\psi_{1xx}\big( \tfrac{1}{6}A_x^3 + \tfrac{1}{2}(\Phi_x^2-1)A^2A_x-A_t \big)
+\psi_{1x} \big( \tfrac{1}{2}(3\Phi_x^2-1) AA_x^2 +\Phi_x\Phi_{xx}A^2A_x \big)
\\&\qquad
-\psi_{1} \big( \tfrac{1}{2}(3\Phi_x^4-2\Phi_x^2-3)A^2A_x -A_t 
+(2\Phi_x^2-1)\Phi_x\Phi_{xx} A^3 \big)
\Big) dx\; dt ,
\end{aligned}
\label{weak-hirota-peakon-eqn-Re}
\\
& \begin{aligned}
0=&\iint_{\Rnum^2} \Big( 
\psi_{2xx} A \Phi_t
+\psi_{2x}\big( \tfrac{5}{6}\Phi_xA_x^3 +\tfrac{1}{2}\Phi_{xx}A A_x^2
+\tfrac{1}{2}(\Phi_x^2 -3)\Phi_xA^2A_x  +\tfrac{1}{2}(\Phi_x^2-1)\Phi_{xx} A^3 \big)
\\&\qquad
+\psi_{2}\big( 2\Phi_x AA_x^2 +\Phi_{xx}A^2A_x 
+(\Phi_x^2+1)(3-\Phi_x^2)\Phi_x A^3
+\Phi_t A \big)
\Big) dx\; dt .
\end{aligned}
\label{weak-hirota-peakon-eqn-Im}
\end{align}
This pair of integral equations \eqref{weak-hirota-peakon-eqn-Re}--\eqref{weak-hirota-peakon-eqn-Im} 
is satisfied by all classical solutions of the Hirota-type peakon equation \eqref{peakon-odd-eqn}. 
\end{lem}

All non-smooth solutions \eqref{polarform} of these integral equations 
will be \emph{weak-amplitude solutions} of the corresponding peakon equations
on the real line, $x\in\Rnum$. 

To derive the 1-peakon solutions of 
the NLS-type peakon equation \eqref{peakon-even-eqn} 
and the Hirota-type peakon equation \eqref{peakon-odd-eqn}, 
we begin by substituting expression \eqref{oscil-peakon} 
in polar form $A=a e^{-|\xi|}$, $\Phi=\phi+\w t+\k x$ 
into the corresponding weak-amplitude integral equations. 
Next, we change the spatial integration variable from $x$ to $\xi$
and split its integration domain into $(-\infty,0)$ and $(0,\infty)$. 
Finally, we use integration by parts to evaluate the integrals. 

\subsection{Hirota peakons}

The Hirota peakon weak-amplitude integrals \eqref{weak-hirota-peakon-eqn-Re}--\eqref{weak-hirota-peakon-eqn-Im} 
are respectively given by
\begin{equation}
\begin{aligned}
0  = & \iint_{\Rnum^2} \Big(
\big( c a e^{-|\xi|} +\tfrac{1}{2}(3\k^4 -2\k^2 -3) a^3 e^{-3|\xi|} \big)\sgn(\xi)\psi_{1}(\xi+ct,t)
\\&\qquad\qquad
+\big( \tfrac{1}{2} (3\k^2-1) a^3 e^{-3|\xi|} \big)\psi_{1x}(\xi+ct,t)
\\&\qquad\qquad
+\big( \tfrac{1}{6} (2-3\k^2) a^3 e^{-3|\xi|} - c a e^{-|\xi|} \big)\sgn(\xi)\psi_{1xx}(\xi+ct,t)
\Big) d\xi\; dt 
\\
= & 
\iint_{\Rnum^2}  \big( \tfrac{1}{2}\k^2(3\k^2 -2) a^3 e^{-3|\xi|} \big)\sgn(\xi)\psi_{1}(\xi+ct,t)\, d\xi\; dt 
\\&\qquad
+a(2c+a^2 \k^2 -\tfrac{2}{3}a^2)\int_{\Rnum} \psi_{1x}(ct,t) \;dt 
\end{aligned}
\end{equation}
and
\begin{equation}
\begin{aligned}
0  = & \iint_{\Rnum^2} \Big(
\big( \w a e^{-|\xi|} -\tfrac{1}{2}\k(\k^4 -2\k^2 -7) a^3 e^{-3|\xi|} \big)\psi_{2}(\xi+ct,t)
\\&\qquad\qquad
+\big( \tfrac{1}{6}\k (3\k^2 -4) a^3 e^{-3|\xi|} \big)\sgn(\xi)\psi_{2x}(\xi+ct,t)
-\big( \w a e^{-|\xi|} \big) \psi_{2xx}(\xi+ct,t)
\Big) d\xi\; dt 
\\
= & 
\iint_{\Rnum^2}  
\big( \tfrac{1}{2}\k(3 +5\k^2 -\k^4)a^3 e^{-3|\xi|} \big)\psi_{2}(\xi+ct,t)
d\xi\; dt 
\\&\qquad
+a(2\w-3a^2\k^3 +\tfrac{4}{3}a^2\k)\int_{\Rnum} \psi_{1}(ct,t)\; dt . 
\end{aligned}
\end{equation}
This pair of equations must hold for all test functions $\psi_{1}$ and $\psi_{2}$, 
respectively.
Hence, we obtain the conditions 
\begin{gather}
a\k(3\k^2 -2) =0,
\quad
a(2c+a^2\k^2 -\tfrac{2}{3}a^2) =0, 
\\
a\k(\k^4 -5\k^2 -3) =0,
\quad
a(2\w-3a^2\k^3 +\tfrac{4}{3}a^2\k) =0 . 
\end{gather}
We want $a\neq 0$, and so then these four conditions yield
\begin{equation}\label{hirota-oscil-peakon-vals}
\k=0,
\quad
\w=0,
\quad
c = \tfrac{1}{3} a^2 . 
\end{equation}
Thus, 
the 1-peakon solution of the Hirota-type peakon equation \eqref{peakon-odd-eqn} 
is given by 
\begin{equation}\label{hirota-oscil-peakon-soln}
u = a \exp(\inum \phi -|x-\tfrac{1}{3}a^2t|),
\quad
a,\phi=\const
\end{equation}
which represents a complex-valued peaked travelling wave with a constant phase. 
In fact, this solution is simply the mCH peakon \eqref{mCH-1peakon}
multiplied by the phase $e^{\inum \phi}$,
and its existence is a direct consequence of 
the Hirota-type peakon equation being $U(1)$-invariant 
and reducing to the mCH equation under $u=\bar u$. 

In contrast, 
the NLS-type peakon equation \eqref{peakon-even-eqn} 
becomes trivial under $u=\bar u$,
and so we expect that its peakon solution will be qualitatively different
than the mCH peakon \eqref{mCH-1peakon} and the Hirota peakon \eqref{hirota-oscil-peakon-soln}.

\subsection{NLS peakon breathers}

The NLS peakon weak-amplitude integrals \eqref{weak-nls-peakon-eqn-Re}--\eqref{weak-nls-peakon-eqn-Im} 
are respectively given by
\begin{equation}
\begin{aligned}
0  = & \iint_{\Rnum^2} \Big(
\big( c a e^{-|\xi|} -\tfrac{1}{6}(21\k^2 +20) a^3 e^{-3|\xi|} \big)\sgn(\xi)\psi_{1}(\xi+ct,t)
\\&\qquad\qquad
-\big( 2\k a^3 e^{-3|\xi|} \big) \psi_{1x}(\xi+ct,t)
+\big( \k a^3 e^{-3|\xi|} - c a e^{-|\xi|} \big)\sgn(\xi)\psi_{1xx}(\xi+ct,t)
\Big) d\xi\; dt 
\\
= & 
-\iint_{\Rnum^2} \big( \tfrac{1}{6}\k(21\k^2+2) a^3 e^{-3|\xi|} \big)\sgn(\xi)\psi_{1}(\xi+ct,t)\, d\xi\; dt 
\\&\qquad
+2a(c-a^2\k)\int_{\Rnum} \psi_{1x}(ct,t) \;dt 
\end{aligned}
\end{equation}
and
\begin{equation}
\begin{aligned}
0  = & \iint_{\Rnum^2} \Big(
\big( \w a e^{-|\xi|} +\tfrac{1}{2}(\k^2+2)(3\k^2 -1) a^3 e^{-3|\xi|} \big)\psi_{2}(\xi+ct,t)
\\&\qquad\qquad
+\big( \tfrac{1}{6}(2-9\k^2) a^3 e^{-3|\xi|} \big)\sgn(\xi)\psi_{2x}(\xi+ct,t)
-\big( w a e^{-|\xi|} \big) \psi_{2xx}(\xi+ct,t)
\Big) d\xi\; dt 
\\
= & 
\iint_{\Rnum^2} \big( \tfrac{1}{2}\k^2(3\k^2-4)a^3 e^{-3|\xi|} \big)\psi_{2}(\xi+ct,t)
d\xi\; dt 
\\&\qquad
+a(2\w +3a^2\k^2 -\tfrac{2}{3}a^2)\int_{\Rnum} \psi_{2}(ct,t) \;dt .
\end{aligned}
\end{equation}
Since this pair of equations must hold for all test functions $\psi_{1}$ and $\psi_{2}$, 
we obtain the respective conditions 
\begin{gather}
a\k(21\k^2 +2) =0, 
\quad
a(c -a^2\k) =0, 
\\
a \k(3\k^2-4) =0, 
\quad
a(2\w +3a^2 \k^2 -\tfrac{2}{3}a^2)  =0 , 
\end{gather}
where we want $a\neq 0$. 
These four conditions yield
\begin{equation}\label{nls-oscil-peakon-vals}
c=0,
\quad
\k=0,
\quad
\w = \tfrac{1}{3} a^2 , 
\end{equation}
and thus the 1-peakon solution of the NLS-type peakon equation \eqref{peakon-even-eqn}
is given by 
\begin{equation}\label{nls-oscil-peakon-soln}
u = a \exp(\inum (\phi+\tfrac{1}{3}a^2 t) -|x|),
\quad
a,\phi=\const.
\end{equation}
This solution represents a \underline{peaked breather}:
it has a stationary peakon profile $|u|= a e^{-|x|}$ 
that is temporally modulated by $e^{\inum \w t}$
where the oscillation frequency is given by 
$\nu = \w/(2\pi) = \tfrac{1}{6\pi}a^2$ 
in terms of the amplitude $a$. 

The existence of this novel type of peakon solution indicates that 
the NLS-type peakon equation \eqref{peakon-even-eqn} 
can be expected to exhibit qualitatively new phenomena 
compared to other peakon equations. 

\subsection{Periodic peakons and peakon-breathers}

Both of the peakon solutions \eqref{hirota-oscil-peakon-soln} and \eqref{nls-oscil-peakon-soln} 
are defined on the real line, $x\in \Rnum$. 
We will next consider spatially periodic counterparts of these solutions,
which are defined on the circle, $x\in S^1$. 

We use the same weak-amplitude formulation stated in Lemma~\ref{lem:weakform}
except that the integration domain $\Rnum^2$ for $(t,x)$ is now replaced by 
$\Rnum\times S^1$ 
and the smooth test functions $\psi_1(t,x),\psi_2(t,x)$ are now periodic in $x$. 
All non-smooth solutions \eqref{polarform} of the resulting integral equations 
will be periodic weak-amplitude solutions on the circle, $x\in S^1$. 
Hereafter we will use the notation $\{\xi\}=\xi-\lfloor\xi\rfloor\in [0,1)$ to denote the fractional part of a variable $\xi$,
where $\lfloor\xi\rfloor$ is the floor function. 

Real periodic peakons are peaked continuous travelling waves on $x\in [0,1]=\Rnum/\Znum$, 
where the endpoints $x=0$ and $x=1$ are identified with each other. 
The general form for a real periodic peakon can be expressed as 
\begin{equation}\label{periodic-peakon}
u = a (\exp(-\{|\xi|\}) + \exp(-\{|\xi-1|\})) 
= a (\exp(-\{\xi\}) + \exp(\{\xi\}-1))
\end{equation}
with $\xi= x-c t$, 
where $a$ is the amplitude and $c$ is the wave speed. 
An equivalent way of writing this expression is given by 
\begin{equation}\label{periodic-peakon-cosh}
u = \tilde a \cosh(\{\xi\} -\tfrac{1}{2})
\end{equation}
where $\tilde a = 2a e^{-\frac{1}{2}}$. 

For both the NLS-type peakon equation \eqref{peakon-even-eqn} 
and the Hirota-type peakon equation \eqref{peakon-odd-eqn},
we will seek oscillatory periodic peakon solutions
\begin{equation}\label{oscil-periodic-peakon}
u = \tilde a \exp(\inum (\phi+\w t+\k \xi)) \cosh(\{\xi\} -\tfrac{1}{2}), 
\quad
\xi= x-c t,
\quad
a,c,\phi,\w,\k=\const
\end{equation}
whose form consists of a real periodic peakon modulated by 
an oscillatory plane-wave phase $\exp(\inum (\phi+\w t+\k x)$. 
We use the same steps as in the derivation of the oscillatory peakon solutions
\eqref{hirota-oscil-peakon-soln} and \eqref{nls-oscil-peakon-soln}. 

The Hirota peakon weak-amplitude integrals \eqref{weak-hirota-peakon-eqn-Re}--\eqref{weak-hirota-peakon-eqn-Im} formulated on $\Rnum\times S^1$ 
are respectively given by
\begin{equation}
\begin{aligned}
0  = & \iint_{\Rnum\times S^1} \Big(
\big( \tfrac{1}{2}(3+2\k^2 -3\k^4) a^3 \cosh(\xi-\tfrac{1}{2})^2 -ca \big) \sinh(\xi-\tfrac{1}{2}) \psi_{1}(\xi+ct,t)
\\&\qquad\qquad
+\big( \tfrac{1}{2} (3\k^2-1) a^3 \cosh(\xi-\tfrac{1}{2}) \big) \sinh(\xi-\tfrac{1}{2})^2 
\psi_{1x}(\xi+ct,t)
\\&\qquad\qquad
+\big( \tfrac{1}{6} (3\k^2-2) a^3 \cosh(\xi-\tfrac{1}{2})^2
+\tfrac{1}{6} (6c-a^2) a \big)\sinh(\xi-\tfrac{1}{2}) \psi_{1xx}(\xi+ct,t)
\Big) d\xi\; dt 
\end{aligned}
\end{equation}
and
\begin{equation}
\begin{aligned}
0  = & \iint_{\Rnum\times S^1} \Big(
\big( \w a -2\k a^3  -\tfrac{1}{2}\k(\k^4 -2\k^2 -7) a^3 \cosh(\xi-\tfrac{1}{2})^2 \big)\cosh(\xi-\tfrac{1}{2}) \psi_{2}(\xi+ct,t)
\\&\qquad\qquad
+\big( \tfrac{5}{6}\k a^3 -\tfrac{1}{6}\k (3\k^2 -4) a^3 \cosh(\xi-\tfrac{1}{2})^2 \big)\sinh(\xi-\tfrac{1}{2})  \psi_{2x}(\xi+ct,t)
\\&\qquad\qquad
-\big( \w a \cosh(\xi-\tfrac{1}{2}) \big) \psi_{2xx}(\xi+ct,t)
\Big) d\xi\; dt .
\end{aligned}
\end{equation}
Using integration by parts and periodicity of the test functions, 
we find that these two integrals yield 
\begin{equation}
\begin{aligned}
0  = & \iint_{\Rnum\times S^1} 
\big( \tfrac{1}{2}\k^2 a^3 -\tfrac{1}{2}\k^2 (3\k^2 -2) a^3 \cosh(\xi-\tfrac{1}{2})^2 \big)\sinh(\xi-\tfrac{1}{2})  \psi_{1}(\xi+ct,t)\; d\xi\; dt 
\\&\qquad
+a\sinh(\tfrac{1}{2})\big( \cosh(\tfrac{1}{2})^2 a^2(\k^2 -\tfrac{2}{3}) -\tfrac{1}{3}a^2 +2c \big) \int_{\Rnum} \psi_{1x}(ct,t) \;dt 
\end{aligned}
\end{equation}
and 
\begin{equation}
\begin{aligned}
0  = & \iint_{\Rnum\times S^1} 
\big( \tfrac{1}{2}\k(3+5\k^2 -\k^4) a^3 \cosh(\xi-\tfrac{1}{2})^2 - \tfrac{1}{2}\k(2\k^2+3) a^3 \big)\cosh(\xi-\tfrac{1}{2}) \psi_{2}(\xi+ct,t) \; d\xi\; dt 
\\&\qquad
-a\sinh(\tfrac{1}{2})\big( \cosh(\tfrac{1}{2})^2 a^2\k(\k^2 -\tfrac{4}{3}) -\tfrac{5}{3}a^2\k -2w\big) \int_{\Rnum} \psi_{2}(ct,t) \;dt .
\end{aligned}
\end{equation}
This pair of equations must hold for all test functions $\psi_{1}$ and $\psi_{2}$,
and hence we obtain the conditions 
\begin{equation}
\begin{gathered}
\k a=0 , 
\quad
\k (3\k^2 -4) a=0 ,
\quad
a\big( \cosh(\tfrac{1}{2})^2 a^2(\k^2 -\tfrac{2}{3}) -\tfrac{1}{3}a^2 +2c \big) =0, 
\\
\k(\k^4 -5\k^2 -3) a =0,
\quad
\k(2\k^2+3) a =0,
\quad
a\big( \cosh(\tfrac{1}{2})^2 a^2(\k^2 -\tfrac{4}{3}) -\tfrac{5}{3}a^2\k -2w\big) =0. 
\end{gathered}
\end{equation}
Since we want $a\neq 0$, these six conditions yield
\begin{equation}
\k=0,
\quad
\w=0,
\quad
c = \tfrac{1}{6}(2\cosh(\tfrac{1}{2})^2+1) a^2= \tfrac{1}{6}(2+\cosh(1)) a^2 .
\end{equation}
Thus, 
the periodic 1-peakon solution of the Hirota-type peakon equation \eqref{peakon-odd-eqn} 
is given by 
\begin{equation}\label{hirota-oscil-periodic-peakon-soln}
u = \tilde a \exp(\inum \phi) \cosh(\{ x-ct\} -\tfrac{1}{2}),
\quad
c = \tfrac{1}{6}(2+\cosh(1)) \tilde a^2 ,
\end{equation}
which has the equivalent form 
\begin{equation}
u = a \exp(\inum \phi)\big(
\exp( -\{|x-\tfrac{1}{12}a^2(e^{-2}+4e^{-1}+1)t|\})
+ \exp( -\{|x-\tfrac{1}{12}a^2(e^{-2}+4e^{-1}+1)t -1|\}) \big) . 
\end{equation}
This solution represents a complex-valued periodic peaked travelling wave 
with a constant phase. 
In fact, the amplitude of the solution is the same as 
the real periodic peakon \cite{QuLiuLiu} of the mCH equation \eqref{mch-eqn},
which is a direct consequence of 
the Hirota-type peakon equation reducing to the mCH equation under $u=\bar u$. 

The NLS peakon weak-amplitude integrals \eqref{weak-nls-peakon-eqn-Re}--\eqref{weak-nls-peakon-eqn-Im} formulated on $\Rnum\times S^1$ 
are respectively given by
\begin{equation}
\begin{aligned}
0  = & \iint_{\Rnum\times S^1} \Big(
\big( \tfrac{1}{6}(21\k+20) a^3 \cosh(\xi-\tfrac{1}{2})^2 -\tfrac{1}{6}(5\k a^2 +6c)a \big) \sinh(\xi-\tfrac{1}{2}) \psi_{1}(\xi+ct,t)
\\&\qquad\qquad
-\big( 2\k a^3 \cosh(\xi-\tfrac{1}{2}) \sinh(\xi-\tfrac{1}{2})^2 \big) 
\psi_{1x}(\xi+ct,t)
\\&\qquad\qquad
+\big( ca - \k a^3 \cosh(\xi-\tfrac{1}{2})^2 \big)\sinh(\xi-\tfrac{1}{2}) \psi_{1xx}(\xi+ct,t)
\Big) d\xi\; dt 
\end{aligned}
\end{equation}
and
\begin{equation}
\begin{aligned}
0  = & \iint_{\Rnum\times S^1} \Big(
\big( \w a +\tfrac{1}{2}(2-3\k^2) a^3  +\tfrac{1}{2}(\k^2+2)(3\k^2-1) a^3 \cosh(\xi-\tfrac{1}{2})^2 \big)\cosh(\xi-\tfrac{1}{2}) \psi_{2}(\xi+ct,t)
\\&\qquad\qquad
-\big( \tfrac{1}{6} (9\k^2 -2) a^3 \cosh(\xi-\tfrac{1}{2})^2 -\tfrac{1}{6} a^3 \big)\sinh(\xi-\tfrac{1}{2})  \psi_{2x}(\xi+ct,t)
\\&\qquad\qquad
-\big( \w a \cosh(\xi-\tfrac{1}{2}) \big) \psi_{2xx}(\xi+ct,t)
\Big) d\xi\; dt .
\end{aligned}
\end{equation}
Integrating by parts and using periodicity of the test functions, 
we find that these two integrals yield 
\begin{equation}
\begin{aligned}
0  = & \iint_{\Rnum\times S^1} 
\big( \tfrac{1}{6}\k (21\k^2 +2) a^3 \cosh(\xi-\tfrac{1}{2})^2 -\tfrac{5}{6}\k a^3 \big)\sinh(\xi-\tfrac{1}{2})  \psi_{1}(\xi+ct,t)\; d\xi\; dt 
\\&\qquad
+2a\sinh(\tfrac{1}{2})\big( \cosh(\tfrac{1}{2})^2 a^2\k +c \big) \int_{\Rnum} \psi_{1x}(ct,t) \;dt 
\end{aligned}
\end{equation}
and 
\begin{equation}
\begin{aligned}
0  = & \iint_{\Rnum\times S^1} 
\big( \tfrac{1}{2}\k^2(3\k^2 -4) a^3 \cosh(\xi-\tfrac{1}{2})^2 + \tfrac{3}{2}\k^2 a^3 \big)\cosh(\xi-\tfrac{1}{2}) \psi_{2}(\xi+ct,t) \; d\xi\; dt 
\\&\qquad
+a\sinh(\tfrac{1}{2})\big( \cosh(\tfrac{1}{2})^2 a^2(3\k^2 -\tfrac{2}{3}) -\tfrac{1}{3}a^2 +2w\big) \int_{\Rnum} \psi_{2}(ct,t) \;dt .
\end{aligned}
\end{equation}
Since this pair of equations must hold for all test functions $\psi_{1}$ and $\psi_{2}$, 
we obtain the respective conditions 
\begin{equation}
\begin{gathered}
\k (21\k^2 +2) a=0,
\quad
\k a=0, 
\quad
a\big( \cosh(\tfrac{1}{2})^2 a^2\k +c \big)=0, 
\\
\k(3\k^2 -4) a=0,
\quad
\k a =0,
\quad
a\big( \cosh(\tfrac{1}{2})^2 a^2(3\k^2 -\tfrac{2}{3}) -\tfrac{1}{3}a^2 +2w\big) =0
\end{gathered}
\end{equation}
where we want $a\neq 0$. 
These six conditions yield
\begin{equation}
c=0,
\quad
\k=0,
\quad
\w = \tfrac{1}{6}(2\cosh(\tfrac{1}{2})^2+1) a^2= \tfrac{1}{6}(2+\cosh(1)) a^2 
\end{equation}
and thus the periodic 1-peakon solution of the NLS-type peakon equation \eqref{peakon-even-eqn}
is given by 
\begin{equation}\label{nls-oscil-periodic-peakon-soln}
u = \tilde a \exp(\inum (\phi+\w t)) \cosh(\{x\} -\tfrac{1}{2}),
\quad
\w = \tfrac{1}{6}(2+\cosh(1)) \tilde a^2 . 
\end{equation}
An equivalent form is 
\begin{equation}
u = a \exp\big(\inum (\phi+\tfrac{1}{12} a^2(e^{-2}+4e^{-1}+1) t)\big)
\big( \exp(-\{x\}) + \exp(\{x\}-1) \big) . 
\end{equation}
This solution represents a \underline{periodic peaked breather}:
it has a stationary periodic peakon profile $|u|= \tilde a\cosh(\{x\}-\tfrac{1}{2})$
that is temporally modulated by $e^{\inum \w t}$.

\section{Concluding remarks}
\label{remarks}

In this paper we have derived two integrable $U(1)$-invariant peakon equations 
from the NLS hierarchy. 
These integrable equations are associated with the first two flows in this hierarchy, 
which consist of the NLS equation and the Hirota equation 
(a $U(1)$-invariant version of mKdV equation), 
so consequently one equation can be viewed as an NLS-type peakon equation 
and the other can be viewed as a Hirota-type peakon equation
(a complex analog of the mCH/FORQ equation). 

For both peakon equations, we have obtained 
a Lax pair, a recursion operator, a bi-Hamiltonian formulation, 
and a hierarchy of symmetries and conservation laws. 
These two peakon equations have been derived previously 
as real 2-component coupled systems \cite{XiaQia,XiaQiaZho} 
by Lax pair methods, without consideration of the NLS hierarchy. 

We have also investigated oscillatory peakon solutions for these $U(1)$-invariant peakon equations. 
The Hirota peakon equation possesses only a moving non-oscillatory peakon
with a constant phase. 
In contrast, 
the NLS peakon equation possesses a stationary oscillatory peakon 
representing a peaked breather. 
Breathers are familiar solutions for soliton equations 
but they had not been found previously for peakon equations. 
We have also obtained spatially periodic counterparts of these peakon solutions.

An important goal will be to find multi-peakon solutions 
for both the NLS-type peakon equation \eqref{peakon-even-eqn} 
and the Hirota-type peakon equation \eqref{peakon-odd-eqn}. 
These solutions should be given by a superposition of oscillatory 1-peakons
\begin{equation}\label{multi-modulated-peakon}
u = \sum_{j=1}^{N} A_j \exp(\inum\Phi_j),
\quad
A_j = \alpha_j(t) \exp(-|x-\beta_j(t)|),
\quad
\Phi_j = \phi_j+\omega_j(t)
\end{equation}
with time-dependent positions $\beta_j(t)$, amplitudes $\alpha_j(t)$, and phases $\omega_j(t)$. 
However, 
the weak-amplitude formulation in Lemma~\ref{lem:weakform}
is not general enough to allow solutions of this form to be found,
and the lack of a more general weak formulation implies that 
some type of explicit regularization of products of distributions 
involving $\sgn(x-\beta_j(t))$ and $\delta(x-\beta_k(t))$, $j,k=1,\ldots,N$, 
will have to be considered. 
This problem will be addressed elsewhere. 

There are several directions that can be pursued for future work:
(1) studying solutions with non-zero asymptotic boundary conditions
and smooth soliton solutions;
(2) establishing existence of weak solutions with general initial data 
and finding conditions for wave breaking;
(3) finding a geometric realization of both peakon equations 
based on the well-known equivalence of the NLS equation to the vortex filament equation.

\section*{Appendix}

\subsection{Scaling formula}

When a Hamiltonian structure possesses a scaling symmetry, 
an explicit formula can be derived to express the Hamiltonian in terms of its gradient,
by using the scaling method in \Ref{Anc03}. 

Consider a Hamiltonian evolution equation 
\begin{equation}\label{H-eqn}
v_t = \H(\delta H/\delta v)
\end{equation}
where 
\begin{equation}\label{H-dens}
H=\int_{-\infty}^{\infty} h(x,v,v_x,\ldots)\,dx
\end{equation}
is a Hamiltonian functional,
and $\H$ is a Hamiltonian operator. 
Note the Hamiltonian gradient is given by 
\begin{equation}\label{H-grad}
Q = \delta H/\delta v = E_v(h)
\end{equation}
where 
\begin{equation}\label{eulerop}
E_v(f)  = \partial_{v} f -D_x(\partial_{v_x} f) +D_x^2(\partial_{v_{xx}} f) + \cdots 
\end{equation}
denotes the Euler operator \cite{Olv-book} with respect to $v$. 

Suppose the evolution equation possesses a scaling symmetry 
\begin{equation}\label{scaling}
x\to \lambda^a x,
\quad
t\to \lambda^b t,
\quad
v\to \lambda^c v
\end{equation}
where $a,b,c=\const$ are the scaling weights. 
Also suppose the Hamiltonian structure is scaling homogeneous
\begin{equation}\label{scaling-h}
h\to \lambda^p h,
\quad
\H\to \lambda^q \H
\end{equation}
where the scaling weights $p,q=\const$ are related to $a,b,c$ by 
\begin{equation}
p+q+b-2c=0
\end{equation}
due to homogeneity. 
Then the Hamiltonian gradient has the scaling 
\begin{equation}
Q\to \lambda^{p-c} Q . 
\end{equation}
Note the generator of the scaling is given by the vector field 
\begin{equation}\label{scal-generator}
\X_{\scal}= (cv -axv_x-btv_t)\partial_v
\end{equation}
in evolutionary form. 

The scaling homogeneity of the Hamiltonian density $h$ implies 
\begin{equation}
\pr\X_{\scal} h = p h -axD_x h - btD_t  h 
\end{equation}
under the prolongation of the scaling generator. 
The explicit action of this generator is given by 
\begin{equation}
\pr\X_{\scal} h \equiv (cv -axv_x-btv_t)E_v(h) = (cv -axv_x-btv_t)Q
\end{equation}
where $\equiv$ denotes equality modulo total $x$-derivatives. 
This yields 
\begin{equation}\label{H-scal}
p h \equiv (cv -axv_x -btv_t)Q +axD_x h +btD_t h
\end{equation}
which can be simplified by the following steps. 

First, the Hamiltonian structure implies \cite{Olv-book} that the Hamiltonian 
is conserved, $\frac{d}{dt}H=0$, 
on the space of solutions $v(t,x)$ of the evolution equation \eqref{H-eqn}. 
This conservation equation can be expressed in the form 
$D_t h = D_x\Phi$ 
for some flux $\Phi$. 
Consequently, in the scaling equation \eqref{H-scal}, 
the term $btD_t h= D_x(bt\Phi)$ is a total $x$-derivative. 

Next, the property that a Hamiltonian operator is skew \cite{Olv-book}
implies that $Q\H(Q) = D_x\Theta$ holds for some expression $\Theta$. 
Hence, on the space of solutions $v(t,x)$ of the evolution equation \eqref{H-eqn}, 
the term $btv_tQ = btQ\H(Q)= bt D_x\Theta = D_x(bt \Theta)$
in the scaling equation \eqref{H-scal} is a total $x$-derivative. 

Finally, the identity $xf = D_x(xD\inv_x f) - D\inv_x f$ 
can be used to express the terms $axD_x h-axv_xQ$ in the scaling equation \eqref{H-scal}
as $ax(D_x h -v_xQ) \equiv aD\inv_x(v_xQ) -ah$. 

Hence, the scaling equation \eqref{H-scal} simplifies to give 
\begin{equation}
(p+a) h \equiv cvQ +aD\inv_x(v_xQ) 
= (c+a)vQ -aD\inv_x(vD_xQ)
\end{equation}
which provides an explicit formula for obtaining the Hamiltonian density 
in terms of its gradient whenever $p+a\neq 0$. 
This condition is equivalent to $\pr\X_{\scal} H \neq 0$. 

\begin{lem}
For any scaling invariant Hamiltonian evolution equation \eqref{H-eqn}, 
the Hamiltonian can be recovered from its gradient by the scaling formula
for the Hamiltonian density (modulo a total $x$-derivative) 
\begin{equation}\label{H-Q-expr}
h \equiv \tfrac{1}{p+a}\big( (c+a)vQ -aD\inv_x(vD_xQ) \big)
\end{equation}
provided that the Hamiltonian is not scaling invariant, $p+a\neq 0$, 
where $a,c,p$ are the scaling weights \eqref{scaling}--\eqref{scaling-h}. 
\end{lem}

\subsection{Helmholtz conditions}

An expression $Q(x,v,v_x,\ldots)$ is a Hamiltonian gradient \eqref{H-grad}
iff it satisfies the Helmholtz conditions \cite{Olv-book}. 
These conditions state that the Frechet derivative of $Q$ with respect to $v$ 
is required to be a self-adjoint operator. 

Let $\delta v$ be an infinitesimal variation of $v$, namely 
$\X v = \delta v$ where $\X= \delta v\partial_v$ is the corresponding vector field. 
Then the Frechet derivative of $Q$ is given by 
\begin{equation}
\delta Q = 
\delta v\partial_v Q +(D_x\delta v)\partial_{v_x} Q +(D_x^2\delta v) \partial_{v_{xx}} Q + \cdots . 
\end{equation}
Its adjoint is defined by integration by parts
\begin{equation}
\delta^* Q= 
\delta  v \partial_v Q -D_x(\delta v\partial_{v_x}Q)) +D_x^2(\delta v\partial_{v_{xx}}Q) + \cdots . 
\end{equation}
Hence, the Frechet derivative is a self-adjoint operator iff 
$\delta Q = \delta^* Q$ holds for an arbitrary $\delta v$. 
Self-adjointness can be expressed in an equivalent form \cite{Olv-book}
by using the higher Euler operators:
\begin{equation}\label{highereulerop}
E_v^{(l)}(Q)  = 
\partial_{\partial_x^l v} Q
-\txtbinom{l+1}{l}D_x\big(\partial_{\partial_x^{l+1} v} Q\big) 
+ \txtbinom{l+2}{l}D_x^2\big(\partial_{\partial_x^{l+2} v} Q\big) 
+ \cdots , 
\quad
l=1,2,\ldots
\end{equation}
where $l=0$ corresponds to the ordinary Euler operator \eqref{eulerop}. 
In terms of the Euler operators \eqref{highereulerop}, 
the Frechet derivative and its adjoint are given by 
\begin{align}
\delta Q& = 
\delta v E_v(Q) + D_x(\delta v E_v^{(1)}(Q)) + D_x^2(\delta v E_v^{(2)}(Q)) + \cdots , 
\\
\delta^* Q & = 
\delta v E_v(Q) -(D_x\delta v) E_v^{(1)}(Q) + (D_x^2\delta v) E_v^{(2)}(Q) + \cdots . 
\end{align}
Hence, $\delta Q = \delta^* Q$ is equivalent to the following explicit conditions. 

\begin{lem}
An expression $Q(x,v,v_x,\ldots)$ is a Hamiltonian gradient \eqref{H-grad}
iff it satisfies 
\begin{subequations}
\begin{align}
\partial_v Q & = E_v(Q)
\\
\partial_{v_x}Q & = -E_v^{(1)}(Q)
\\
\partial_{v_{xx}}Q & = E_v^{(2)}(Q)
\\
& \vdots
\nonumber
\end{align}
\end{subequations}
which comprise the Helmholtz conditions. 
\end{lem}

\section*{Acknowledgements}

S.C.A.\ is supported by an NSERC research grant. 
The reviewers are thanked for helpful comments which have improved this paper.

\end{document}